\documentclass[prl,aps]{revtex4}
\usepackage{graphicx}
\usepackage{amsfonts}
\usepackage{amssymb}
\usepackage{amsmath}
\usepackage{psfrag}
\usepackage{natbib}
\usepackage{mathtools}
\usepackage{subfigure,float}
\usepackage[colorlinks=true, linkcolor=blue, urlcolor=blue, citecolor=black]{hyperref}
\newcommand{\beqa}{\begin{eqnarray}}
\newcommand{\eeqa}{\end{eqnarray}}
\newcommand{\beq}{\begin{equation}}
\newcommand{\eeq}{\end{equation}}

\begin{document}
\title{On the singular nature of the elastocapillary ridge}

\author{A. Pandey$^{1,2}$, B. Andreotti$^3$, S. Karpitschka$^4$, G. J. van Zwieten$^5$, E. H. van Brummelen$^6$, J. H. Snoeijer$^1$
}
\affiliation{
$^1$Physics of Fluids Group, Faculty of Science and Technology,
Mesa+ Institute, University of Twente, 7500 AE Enschede, The Netherlands \\
$^2$ Department of Biological and Environmental Engineering, Cornell University, Ithaca, NY 14853, USA\\
$^3$ Laboratoire de Physique de l'ENS, UMR 8550 Ecole Normale Sup\'erieure -- CNRS -- Universit{\'e}~de~Paris -- Sorbonne~Universit{\'e}, 24 rue Lhomond, 75005 Paris\\
$^4$ Max Planck Institute for Dynamics and Self-Organization (MPIDS), 37077 G\"ottingen, Germany\\
$^5$ Evalf Computing, Burgwal 45, 2611 GG Delft, The Netherlands\\
$^6$ Multiscale Engineering Fluid Dynamics Group, Department of Mechanical Engineering, Eindhoven University of Technology, P.O. Box 513, 5600 MB Eindhoven, The Netherlands
}

\begin{abstract}
The functionality of soft interfaces is crucial to many applications in biology and surface science. Recent studies have used  liquid drops to probe the surface mechanics of elastomeric networks. Experiments suggest an intricate surface elasticity, also known as the Shuttleworth effect, where surface tension is not constant but depends on substrate deformation. However, interpretations have remained controversial due to singular elastic deformations, induced exactly at the point where the droplet pulls the network. Here we reveal the nature of the elastocapillary singularity on a hyperelastic substrate with various constitutive relations for the interfacial energy. First, we finely resolve the vicinity of the singularity using goal-adaptive finite element simulations. This confirms the universal validity, also at large elastic deformations, of the previously disputed Neumann's law for the contact angles. Subsequently, we derive exact solutions of nonlinear elasticity that describe the singularity analytically. These solutions are in perfect agreement with numerics, and show that the stretch at the contact line, as previously measured experimentally, consistently points to a strong Shuttleworth effect. Finally, using Noether's theorem we provide a quantitative link between wetting hysteresis and Eshelby-like forces, and thereby offer a complete framework for soft wetting in the presence of the Shuttleworth effect. 
\end{abstract}
\maketitle

\section{Introduction}
The wetting and adhesion of soft materials have recently become a quickly expanding domain and finds applications in the design of innovative materials (adhesives~\cite{Li378}, slippery surfaces~\cite{Newton2011}, highly stretchable electronics~\cite{Rogers1603}), to analyse the mechanics of cells and biological tissues~\cite{Discher2005aa,Douezan2012aa}, and in between, in the field of bioengineering (reversible adhesives~\cite{King2014}, e-skin~\cite{Zoueaaq0508}, etc). Reticulated polymer networks are model soft materials, with versatile properties. At small length and time scales their structure is liquid-like and highly deformable. At large scales, however, the presence of crosslinks give the polymer networks a finite shear modulus $G$ such that they behave like elastic solids~\cite{binder2011glassy,de1979scaling,doi1988theory,rubinstein2003polymer}. The elasticity is of entropic origin, and as a consequence the elastic moduli of polymer networks can be exceedingly small compared to those of (poly)crystalline materials, whose elasticity is of enthalpic origin.

This dual liquid-solid character of polymer networks has recently led to a strong controversy on the so-called Shuttleworth effect~\cite{Shut50,AS16,SJHD2017,AS2020}, which describes the capillary forces at an elastic interface. The key question is whether  the surface energy $\gamma$ of a soft solid, which is a nano-scale quantity, depends on the amount of stretching, i.e. on the macroscopically applied deformation. If such a dependency exists, then the excess force per unit length in the interfacial region of the solid, which is by definition the surface tension $\Upsilon$, is not equal to the excess energy per unit surface area $\gamma$. The two quantities are related by the Shuttleworth relation~\cite{Shut50},  

\begin{equation}\label{eq:shworth}
\Upsilon=\gamma + \lambda \frac{d \gamma}{d \lambda},
\end{equation}
where $\lambda$ is the stretch of a surface element. This offers an exciting perspective analogous to surface rheology, where surface tension $\Upsilon(\lambda)$ depends on the state of the system -- potentially leading to stiffening or even softening of the interface. However, given that interfacial properties are determined at the nanoscale, the emergence of a Shuttleworth effect for soft polymeric networks is debated \cite{AS2020,Marchand2012c,Bostwick:2014aa,XuNatComm2017,xu2018,Schulman2018aa,Snoeijer2018,liang2018surface,Wu2018,MasurelPRL2019,ChenDanielsSM2019,Gorcum2020}. To a large extent, the discussion is due to a lack of a consistent analytical theory to interpret macroscopic experiments.  

\begin{figure}[t]
    \centering
    \includegraphics[width=1.0\textwidth]{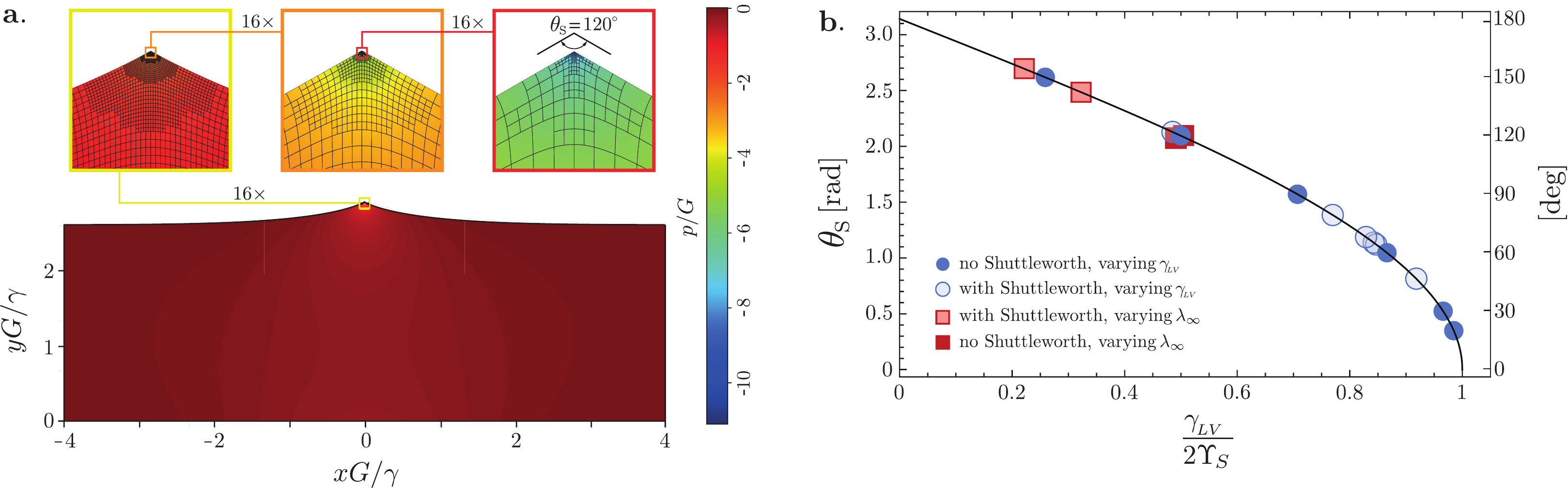}
    \caption{Symmetric wetting ridges under large deformation, with and without Shuttleworth effect. (a) Typical numerical solution, where successive magnifications show the adaptive resolution of the elastocapillary ridge. The example is a case without Shuttleworth effect, with equal liquid and solid surface energies $\gamma$ (giving a solid angle $\theta_S=120^\circ$). The scales are expressed in the corresponding elastocapillary length $\gamma/G$. (b) The solid angle $\theta_S$ versus the ratio of liquid-vapor surface tension $\gamma_{LV}$ and solid surface tension $\Upsilon_S$. Symbols are numerical results with Shuttleworth effect (open symbols, $\Upsilon_S$ measured at the contact line) and without Shuttleworth effect (closed symbols). We varied both $\gamma_{LV}$ (circles), and the amount of prestretch of the substrate from $\lambda_\infty=1$ to 2 (squares). The solid line corresponds to Neumann's law (\ref{eq:neumsymmetric}), with $\Upsilon_S$ based on its value at the contact line.}
           \label{fig:zoom}
\end{figure}

Hitherto, all observations on the Shuttleworth effect in polymer networks are based on ``Soft Wetting"~\cite{AS2020}, where a liquid is partially wetting the substrate. A drop of liquid sitting on a soft amorphous polymeric solid exhibits a shape that is globally similar to that on a non-deformable crystalline solid. However, intermolecular forces are able to deform the soft solid over a scale set by the balance between capillarity and elasticity, known as the elastocapillary length~\cite{Rusanov:1975aa,Shanahan1987aa,Carre1996a,White:2003aa,PC2008aa,Jerison2011a,PARKNATURE}. Below this length scale, the soft substrate takes the shape of a sharp ridge that is characterised by the solid angle at its tip. A fundamental question is then how the contact angles, the prime characteristics of wetting, are selected in the hybrid case where both capillarity and elasticity play a role. Is the liquid contact angle with respect to the undeformed substrate still selected by the Young's law? Is the local structure of the interfaces at the contact line selected by a simple force balance, leading to a generalised Neumann's law? What is the role of contact line pinning? The controversies on the existence, or not, of the Shuttleworth effect in soft solids revolve around these questions. For example, recent experiments probing a strain-dependent surface tension~\cite{XuNatComm2017,xu2018} have been based on the measurement of the angle $\theta_S$ made by the solid below the contact line ($\theta_S$ defined in Fig.~\ref{fig:zoom}a). Indeed, such an angle receives a simple explanation when a Neumann force balance of surface tensions is assumed -- as was originally derived using the small deformation theory of linear elasticity~\cite{Marchand2012a,Style2012a,Limat2012a,Style2013b}. However, this interpretation has been challenged by molecular \cite{liang2018surface} and continuum simulations \cite{Wu2018,MasurelPRL2019}, suggesting that the elastic stress contributes to the force balance at the contact line -- potentially giving a change in $\theta_S$ without invoking any Shuttleworth effect. A recent proposal is that the wetting ridge below the contact line could behave like a disclination defect in crystalline solid~\cite{MasurelPRL2019}: in the regime of large deformations, a singular Eshelby force could then emerge at the contact line, which would be involved in the force balance and invalidates the Neumann's law. Numerical simulations using a finite element method may appear to suggest such alternative description of the soft wetting problem \cite{Wu2018,MasurelPRL2019}, where no Shuttleworth effect is present but an elastic singularity appears at the contact line. However, no closed form analytical theory is available to predict the properties of wetting ridges at large deformations \cite{Brummelen:2017sh}.
 
Before trying to analyse the microscopic origin of a potential Shuttleworth effect, implying a strain-dependent surface tension $\Upsilon(\lambda)$, there is an urgent need to clarify the mechanical consequences of the existence of such an effect. In particular, numerical simulations ultimately rely on a mechanical description which must be totally self-consistent, including the possibility of singularities. If such singularities do exist, then non-adaptive numerical approximations become unreliable to obtain the correct solution of a problem. 

In this paper we numerically resolve the problem of soft wetting, using an adaptive numerical technique that allows us to resolve the elastocapillary wetting ridge on all scales (Fig.~\ref{fig:zoom}a). This includes the possibility of singularities, large elastic deformations and the Shuttleworth effect. It is found that the elastic singularity at the wetting ridge is not sufficiently strong to interfere with the balance of surface tensions at the contact line, so that Neumann's law is universally valid -- irrespective of the presence of large deformations, Shuttleworth effect and pinning.  Subsequently, we derive exact solutions to nonlinear elasticity that analytically resolve the ridge-singularity in the presence of large deformations. These asymptotic solutions, valid near the singularity, are fully confirmed by the numerical results and offer an novel route to interpret experiments, via the surface stretch measured at the contact line. Applying our analysis to the strain measurements in~\cite{XuNatComm2017}, we provide further evidence for a strong Shuttleworth effect. Finally, we show how Eshelby-like forces can emerge when the substrate has true defects that represent pinning sites on the substrate, and reveal their effect on the contact angles.  
 
\section{Free energy formulation}

In experiments, the drop size is usually large compared to the elastocapillary length, $\gamma/G$, where $\gamma$ is a typical surface energy (of the solid or the liquid), while $G$ is the shear modulus of the substrate. In this regime, the curvature of the contact line is negligible compared the size of the wetting ridge, and the geometry is quasi-two-dimensional. Below, we therefore formulate the problem in a plane strain description and explain the numerical method that is used to adaptatively resolve the singular nature of the elastocapillary ridge.

\subsection{Minimising the elastocapillary energy}

The statics of wetting amounts to finding the state of minimal elastocapillary energy. The substrate deformation is described by a mapping from the reference state prior to deformation, to a current state after deformation. Following standard notation, the mapping is written as

\begin{equation}
\mathbf x = \chi(\mathbf X),
\end{equation}
where $\mathbf X$ is the position of a material point on the reference domain, mapped onto its current position $\mathbf x$. We consider the geometry to be invariant along the contact line, so that the problem is two-dimensional (plane strain elasticity). Hyperelastic solids are described by an elastic energy density $W(\mathbf F)$, which depends on the deformation gradient tensor $\mathbf F = \partial \mathbf x/\partial \mathbf X$. We now turn to the interface, which in the (plane strain) two-dimensional description is one-dimensional. We define the arc-length material coordinate at the interface as $S$, and the current surface position $\mathbf x_s(S)=\chi(\mathbf X(S))$. The surface stretch, accounting for the change of length of surface elements follows as

\begin{equation}
\lambda^2 = \frac{\partial \mathbf x_s}{\partial S} \cdot  \frac{\partial \mathbf x_s}{\partial S},
\end{equation}
which is a scalar in this plane strain description. Now we can express capillarity, usually defined by the excess energy $\gamma$ per unit area of the \emph{deformed} state, as a free energy $\lambda\gamma$ per unit area in the \emph{reference} state.

Crucially, elastic media can exhibit a nontrivial capillarity where the surface energy $\gamma(\lambda)$ will itself be a function of the stretch $\lambda$ -- this is the Shuttleworth effect~\cite{Shut50,AS16,SJHD2017,AS2020}.
With this, the elastocapillary energy (per unit length along the contact line) takes the form 

\begin{eqnarray}
\label{eq:Echi}
\mathcal E[\chi] 
= \int d^2X \, W(\mathbf F) + \oint dS \, \lambda \gamma(\lambda),
\end{eqnarray}
respectively giving the total (bulk) elastic energy and the (surface) capillary energy. $\mathbf F$ and $\lambda$ are the corresponding bulk and surface stretches, and are both defined by the map $\chi(\mathbf X)$. We anticipate that we will consider incompressible substrates, in which case the constraint of incompressibility will be included in $W(\mathbf F)$.

Equilibrium configurations of the elastocapillary substrate are found by minimising the functional $\mathcal E$ with respect to $\chi(\mathbf X)$. Considering variations $\delta \mathbf x = \delta \chi(\mathbf X)$, we find

\begin{eqnarray}\label{eq:var}
\delta \mathcal E &=&
 \int d^2X \, \left( \frac{\partial W}{\partial \mathbf F} : \delta \mathbf F \right)
 + \oint dS \, \frac{d(\lambda \gamma)}{d\lambda} \delta \lambda  
=\int d^2X \, \left(\mathbf s : \mathrm{Grad}(\delta \mathbf x)  \right)
 + \oint dS \, \left(\Upsilon \mathbf t  \cdot \frac{\partial \delta \mathbf x}{\partial S}\right).
\end{eqnarray}
Here we introduced the nominal (or first Piola-Kirchhoff) stress tensor $\mathbf s$, and the surface tension $\Upsilon$, 

\begin{equation}\label{eq:upsilondef}
\mathbf s = \frac{\partial W}{\partial \mathbf F}, \quad \quad \Upsilon = \frac{d(\lambda \gamma)}{d\lambda} = \gamma + \lambda \frac{d\gamma}{d\lambda},
\end{equation}
where for the latter we indeed recognise the Shuttleworth relation (\ref{eq:shworth}). 
In addition, we used that $\delta \lambda = \mathbf t \cdot \partial \delta \mathbf x/\partial S$ along the boundary, where $\mathbf t$ is the surface-tangent unit vector in the current configuration. 

To study the elastocapillary ridge, we still need to include the pull of the contact line, induced by the liquid drop that is wetting the solid. This can be achieved by making explicit the capillary energy of the drop, via its liquid-vapour surface energy $\gamma_{LV}$. The subtlety here is that one needs to impose a constraint at the contact line~~\cite{LubbersJFM14,Snoeijer2018}: the position $\mathbf x$ of the liquid-vapour interface must (by definition) coincide with that of the solid interface. The effect of this constraint, imposed by a Lagrange multiplier, provides a localised traction on the substrate, pulling with a strength $\gamma_{LV}$ along the direction of the liquid-vapour interface $\mathbf t_{LV}$ \footnote{Note that in the present work the solid interface is treated as part of the substrate, so that the external traction of the liquid-vapour interface is $\gamma_{LV}\mathbf t_{LV}$. The force transmitted onto the elastic \emph{bulk} of the substrate, i.e. after passing the interface, is more intricate as discussed e.g. in~\cite{Marchand2012c,Bostwick:2014aa,AS16}}. The representation by a local force is indeed commonly used in modelling approaches~\cite{Style2012a,Limat2012a,Style2013b,Wu2018,MasurelPRL2019}. Here we therefore treat the contact line as a perfectly localized external traction, with the associated work functional $\mathcal R = \gamma_{LV} \mathbf t_{LV} \cdot  \mathbf x(\mathbf X_{\rm cl})$, where $\mathbf X_{\rm cl}$ is the solid's material point at which the contact line is acting.  During the variation this corresponds to a work
\begin{equation}\label{eq:forcing}
\delta \mathcal R = \gamma_{LV} \mathbf t_{LV} \cdot \delta \mathbf x(\mathbf X_{\rm cl}),
\end{equation}
The virtual work principle, $\delta \mathcal E = \delta \mathcal R$, then gives the equilibrium condition

\begin{eqnarray}\label{eq:weak}
\int d^2X \, \left(\mathbf s : \mathrm{Grad}(\delta \mathbf x)  \right)
 + \oint dS \, \left(\Upsilon \mathbf t  \cdot \frac{\partial \delta \mathbf x}{\partial S}\right) =  
 \gamma_{LV} \mathbf t_{LV} \cdot \delta \mathbf x(\mathbf X_{\rm cl}),
 \end{eqnarray}
 which should be satisfied for arbitrary $\delta \mathbf x$.

Equation~(\ref{eq:weak}) defines the elastocapillary equilibrium in the weak formulation. This equilibrium is indeed highly singular. Namely, the forcing on the right hand side appears as a point force, pulling at $\mathbf X_{\rm cl}$, while the elastocapillary energies on the left contains only surface and bulk contributions. The debate in the literature precisely revolves around the following question: Do singularities appear in surface (capillarity) or in bulk (elasticity), in order to balance the point force at the contact line? 

\subsection{Numerical method}

Our interest pertains to finding equilibrium configurations of the elastocapillary problem, i.e. to minimisers of the energy functional in~(\ref{eq:Echi}) extended with the work functional $\mathcal R$ representing the contact line, subject to appropriate boundary conditions. 
Specifically, we consider substrates that are flat in the reference configuration, with complete fixation at the bottom boundary and guided fixation (slip) at the lateral boundaries. We allow for the possibility to impose a prestretch $\lambda_\infty$, refering to the uniaxial stretch far away from the contact line. Besides the work associated to the point-forcing at the contact line, the top surface is free of traction, as is made explicit in the weak formulation (\ref{eq:weak}) of the minimisation problem. The constitutive relations for the strain-energy density and the surface energy are specified in Section~\ref{sec:FEMresults} below. In all simulations, the shear modulus $G$ and the relevant surface energies are chosen such that the wetting ridge is much smaller than the width of the domain, with a typical example given in Fig.~\ref{fig:zoom}(a). In that example the domain width and height respectively are $8 \gamma_{LV}/G$ and height $\frac{8}{3}\gamma_{LV}/G$, which are representative for all presented results.

Here we numerically approximate the minimiser of $\mathcal E - \mathcal R$ by means of a \emph{goal-adaptive finite-element method}~\cite{BeckerRannacher2001,Oden:2001ss}. In goal-adaptive methods, the finite-element approximation is locally refined on the basis of an a-posteriori error estimate, in such a manner that an optimal approximation to a predefined quantity of interest (the \emph{goal\/}) is obtained. Goal-adaptive finite-element methods generally proceed according to the SEMR (\texttt{Solve} $\rightarrow$ \texttt{Estimate} $\rightarrow$ \texttt{Mark} $\rightarrow$ \texttt{Refine}) process~\cite{Nochetto:2012hl,Brummelen:2017rr}. The SEMR process starts by solving a finite-element approximation on a coarse mesh. Next, the contribution of each element to the error in the goal quantity is estimated, based on a so-called dual problem~\cite{BeckerRannacher2001,Oden:2001ss,Brummelen:2017rr}.
The elements that yield the largest contribution to the error are marked according to a refinement strategy. These marked elements are subsequently refined by subdivision. This process is repeated until a certain threshold for the error estimate is satisfied or a preset number of refinement iterations has been executed. In accordance with our interest in minimisers of $\mathcal E - \mathcal R$, we take the energy itself as the goal functional. The optimality conditions are resolved by means of the Newton--Raphson method. The goal-adaptive finite-element method for the present problem has been implemented in the open-source software framework Nutils~\cite{nutils}. The optimality conditions~(\ref{eq:weak}) are in fact directly derived from an implementation of the energy functional $\mathcal E - \mathcal R$ via the automatic-differentiation functionality in Nutils. 

An illustration of a goal-adaptive finite-element approximation is provided in Fig.~\ref{fig:zoom}(a). The approximation is based on 16 refinement iterations. Accordingly, the smallest elements in the adaptive approximation are $2^{16}$ times smaller than the initial element size. The initial mesh comprises $24\times8$ uniform quadrilateral elements and, correspondingly, the smallest elements are 5--6 orders of magnitude smaller than the elastocapillary length. Importantly, the adaptive procedure automatically introduces the local refinements in the vicinity of the contact line. This refinement pattern is in agreement with the singularity of the pressure towards the contact line, and we extensively verified the numerical convergence of the result. For the result shown in Fig.~\ref{fig:zoom}(a), the relative numerical error in the computed value of the solid opening angle $\theta_S$ is less than $10^{-6}$.

%

\section{Elastocapillary ridges, with and without Shuttleworth effect}\label{sec:FEMresults}


We now present the adaptively resolved numerical results for the elastocapillary ridge. We will consider cases with constant surface energy and with variable surface energy, i.e. without and with Shuttleworth effect. For the bulk elasticity, we will consider materials with a neo-Hookean strain-energy density (using plain strain), 

\begin{equation}\label{eq:Wneohookean}
W(\mathbf F) = \frac{1}{2}G \left( \mathbf F^T \!\!:\! \mathbf F - 2 \right) - p \left( \det \mathbf F - 1\right),
\end{equation}
where we introduced the pressure $p$ to impose the constraint of incompressibility. In contrast to bulk elasticity, there are no standard constitutive relations for the surface energy of soft solids. Here, we propose a surface energy of the form
\begin{equation}\label{eq:rheology}
\gamma_S(\lambda) = \gamma_0 \left( 1 - c_0 \log \lambda + c_1 (\lambda-1) \right).
\end{equation} 
We from now on add the subscript ``$S$" to indicate that we refer to the solid interface (to distinguish from the liquid-vapour surface energy $\gamma_{LV}$). Expanding (\ref{eq:rheology}) around $\lambda=1$ up to quadratic order, one recovers the Ansatz for surface elasticity as proposed in \cite{Gorcum2020}, while if in addition $c_0=c_1$ one finds a linear surface elasticity as proposed in \cite{xu2018}. An advantage of the constitutive relation (\ref{eq:rheology}) is that the logarithm conveniently keeps the system away from $\lambda \rightarrow 0$. The parameters $c_{0,1}$ must satisfy an admissibility condition such that the surface energy remains convex and that both the energy $\gamma_S$ and the surface tension $\Upsilon_S$ remain positive definite. According to the Shuttleworth relation of (\ref{eq:upsilondef}), the above surface energy gives a surface tension
\begin{equation}
\Upsilon_S(\lambda) = \gamma_0 \left( 1 + c_1- c_0 - c_0 \log \lambda + 2 c_1 (\lambda-1) \right),
\end{equation}
and one verifies that ensuring $\Upsilon_S>0$ is sufficient for the constants $c_{0,1}$ to be admissible. Below we present results for the case where $c_{0,1}=0$ (no Shuttleworth effect), and for $c_{0,1}=1$ (strong Shuttleworth effect), which are indeed in the admissible regime. For later reference we also define the associated ``chemical potential"
\begin{equation}\label{eq:gammamu}
\mu_S(\lambda) \equiv \lambda^2 \frac{d\gamma_S}{d\lambda}=  \gamma_0 \left( c_1 \lambda^2 - c_0 \lambda \right),
\end{equation}
which will be relevant in Sec.~\ref{sec:pinning}.

In general, the solid-liquid and liquid-vapour interfaces of course exhibit a different surface constitutive relation, respectively, which we write $\gamma_{SL}(\lambda)$ and $\gamma_{SV}(\lambda)$. For most of the paper we focus on cases where the solid-liquid and solid-vapour energies are identical, and simply denoted $\gamma_{S}(\lambda)$. This renders the problem symmetric around the contact line, so that the equilibrium contact angle of the liquid is $90^\circ$ and the associated forcing is vertical. Also, this symmetry replaces the ``second boundary condition" discussed in~\cite{Snoeijer2018,AS2020}. Asymmetric surface energies will be considered in Sec.~\ref{sec:pinning}, where we adress the relation between pinning, the contact angle, and the second boundary condition.

\subsection{Universality of Neumann's law}
 
We first consider the solid angle $\theta_S$, as measured at the tip of the wetting ridge in FEM. Figure~\ref{fig:zoom}(b) shows $\theta_S$ plotted against $\gamma_{LV}/\Upsilon_S$, with the value of $\Upsilon_S$ taken at the tip. Clearly, $\theta_S$ follows a universal curve for all cases considered. The parameters that were varied in our simulations are the contact line force $\gamma_{LV}$ (compared to the value of $\gamma_0$ in (\ref{eq:rheology})), while solid surface tensions are with or without Shuttleworth effect ($c_{0,1}=0$, respectively $c_{0,1}=1$). We also considered different amounts of prestretch of the substrate, ranging from $\lambda_\infty=1$ (no prestretch) to $\lambda_\infty=2$ (extending the length by 100\%). The universal curve for $\theta_S$ indeed follows Neumann's law, which for the specific case of identical solid-liquid and solid-vapour energies reads
\begin{equation}\label{eq:neumsymmetric}
2\Upsilon_S \sin \left( \frac{1}{2}(\pi - \theta_S) \right) = \gamma_{LV}.
\end{equation}
Here, we emphasise that owing to the Shuttleworth effect, the surface tension $\Upsilon_S(\lambda)$ depends on the strain. Since the Neumann balance is to be interpreted as a boundary condition at the contact line, we consider (\ref{eq:neumsymmetric}) with values of the stretch $\lambda_{\rm cl}$ taken at the contact line. The result of (\ref{eq:neumsymmetric}) is superimposed as the solid line in Fig.~\ref{fig:zoom}(b), providing a perfect description of the FEM results.
 \begin{figure}[t]
    \centering
    \includegraphics[width=0.8\textwidth]{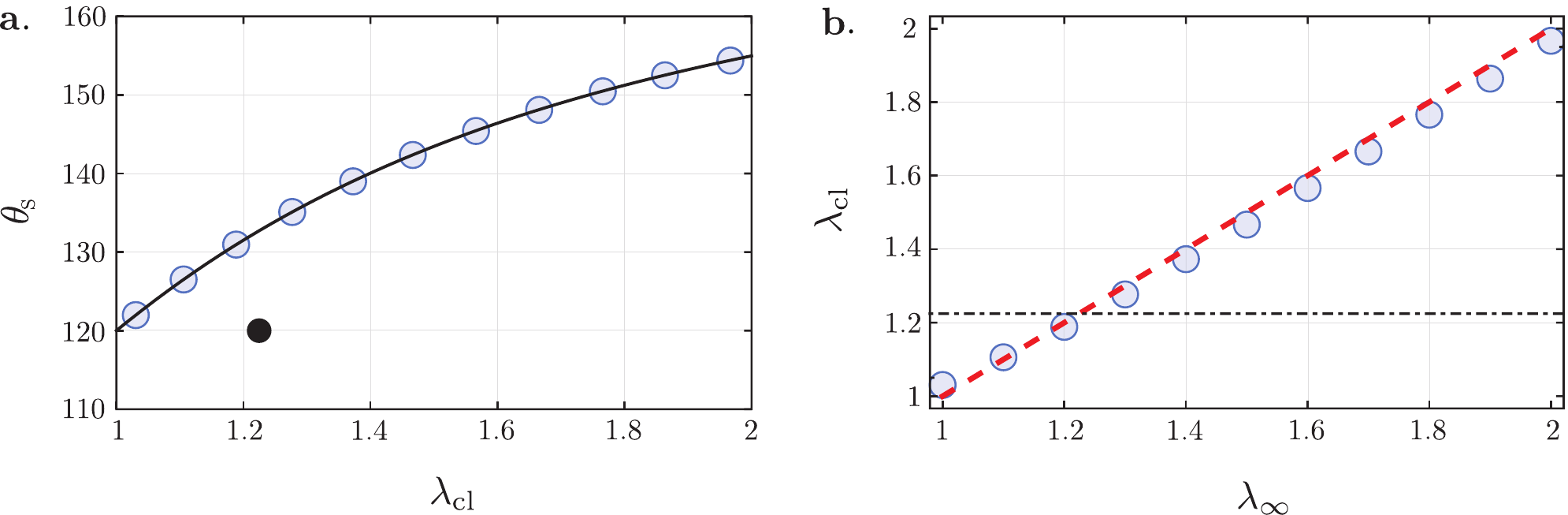}
    \caption{Geometry of the elastocapillary ridge upon stretching the substrate. (a) Solid angle $\theta_S$ as a function of the  stretch at the contact line $\lambda_{\rm cl}$. Open circles correspond to FEM in the presence of a strong Shuttleworth effect ($c_0=c_1=1$, with $\gamma_0=\gamma_{LV}$). Solid line is the analytical prediction by Neumann's law (\ref{eq:neumsymmetric}). The closed circles (several measurements superimposed) corresponds to FEM without a Shuttleworth effect ($c_0=c_1=0$, with $\gamma_0=\gamma_{LV}$). (b) Relation between the stretch at the contact line $\lambda_{\rm cl}$ and the globally imposed stretch $\lambda_\infty$. In the presence of a strong Shuttleworth effect, the two stretches takes on very similar values (red dashed line, $\lambda_{\rm cl}=\lambda_\infty$, is a guide to the eye). Without Shuttleworth effect $\lambda_{\rm cl}=\sqrt{\pi/\theta_S}$ takes on a constant value (dash-dotted line).}
           \label{fig:thetavslambda}
\end{figure}

We thus reach a first major conclusion: Neumann's law (based on the local values of the surface tension) universally applies to elastocapillary wetting ridges, irrespective of the large elastic deformations at the contact line. This rejects the recent hypothesis that strong elastic nonlinearity, as encountered for narrow $\theta_S$ and large prestretch, would lead to a failure of Neumann's law~\cite{MasurelPRL2019}. The universal validity of Neumann's law has an immediate consequence for measurements of the surface-constitutive relation based on $\theta_S$, since we safely conclude that $\theta_S$ gives a direct access to the values of $\Upsilon_S$. Phrased differently, the experimental observation for PDMS that $\theta_S$ increases with $\lambda_\infty$~\cite{XuNatComm2017} can, in a macroscopic theory based on hyperelasticity, only be explained via a strong Shuttleworth effect. 

To further illustrate this, we closely follow the experimental protocol of \cite{XuNatComm2017} in our simulations, and consider how the geometry of the ridge evolves when stretching the substrate by an increasing amount $\lambda_\infty$. Figure~\ref{fig:thetavslambda}(a) shows $\theta_S$ versus the stretch at the contact line $\lambda_{\rm cl}$. The open circles are FEM results with a Shuttleworth effect ($c_{0,1}=1$, and $\gamma_0=\gamma_{LV}$), showing an increase of the solid opening angle $\theta_S$. Indeed, the  dependence of $\theta_S$ is perfectly predicted by Neumann's law  (\ref{eq:neumsymmetric}), as is indicated by the solid line. In experiments, one of course does not control the stretch at the contact line $\lambda_{\rm cl}$, but rather the global stretch of the substrate $\lambda_\infty$. In Fig.~\ref{fig:thetavslambda}(b) we therefore plot these two stretches against one another. While $\lambda_{\rm cl}$ is not exactly identical to the imposed stretched $\lambda_\infty$, the differences turn out to be minor -- consistently with experiments \cite{XuNatComm2017}). As a guide to the eye, the dashed line in Fig.~\ref{fig:thetavslambda}(b) indicates $\lambda_{\rm cl}=\lambda_\infty$. We expect this near-homogeneity of $\lambda$'s to arise only for nearly symmetric $\gamma_{SL}$ and $\gamma_{SV}$, as asymmetry in general leads to stronger gradients of stretch (cf. Sec~\ref{sec:pinning}).

The scenario changes dramatically when the substrate does \emph{not} exhibit a Shuttleworth effect (i.e. $c_{0,1}=0$). In that case, both $\theta_S$ and $\lambda_{\rm cl}$ take on a constant value, that is totally independent of the imposed $\lambda_\infty$. This is indicated in Fig.~\ref{fig:thetavslambda}(a) by the closed circle -- which in fact corresponds to various simulations with $\lambda_\infty$ ranging from 1 to 2. This invariance of $\theta_S$ with respect to $\lambda_\infty$ is easily understood from the Neumann balance. Namely, surface tensions are constant when $c_{0,1}=0$, and since we consider $\gamma_0=\gamma_{LV}$ we find that $\theta_S=120^\circ$. By contrast, the invariance of the stretch at the tip comes as a surprise and its explanation calls for a better understanding the nature of the elastic singularity. Below, we will derive analytically that without the Shuttleworth effect, $\lambda_{\rm cl}=\sqrt{\pi/\theta_S}$, irrespective of the externally imposed prestretch $\lambda_\infty$ of the substrate.

Measurements of the stretch at the contact line thus provide important additional information on the Shuttleworth effect, that till date has not yet been explored. Namely, experiments by \cite{XuNatComm2017} reveal an increase of stretch at the contact line upon a global stretching of the substrate. From the above it is clear that such a dependence can, in a macroscopic theory based on hyperelasticity, not occur when there is no Shuttleworth effect.

\subsection{Stress singularity and the elastic Marangoni effect}

\begin{figure}[t]
    \centering
   \includegraphics[width=1.0\textwidth]{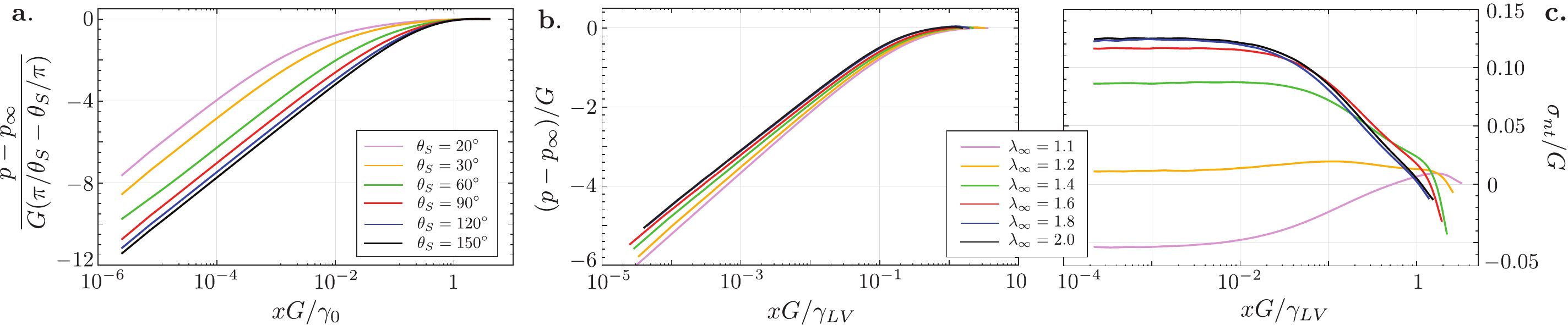}
    \caption{Elastic stress along the free surface near the ridge singularity (symmetric ridges).  (a) Pressure $p$ vs distance to the contact line $x$, scaled as indicated on the axes. Data correspond to the situation without Shuttleworth ($c_{0,1}=0$) with different $\theta_S$ obtained by varying the ratio $\gamma_{LV}/\gamma_0$. (b) Pressure $p$ vs distance to the contact line $x$, scaled as indicated on the axes. Data correspond to the situation with Shuttleworth ($c_{0,1}=1$) with different amounts of prestretch $\lambda_\infty$. (c) Shear stress $\sigma_{nt}$ vs distance to the contact line $x$, scaled as indicated on the axes. Data correspond to the same case as in (b).}
           \label{fig:pressure}
\end{figure}

To further analyse the vicinity of the tip, we now turn to the elastic stress measured along the free surface. In Fig.~\ref{fig:pressure}(a,b) we plot the pressure $p$ as a function of the distance to the contact line $x$, on a semilogarithmic scale. In all cases the FEM simulations exhibit a weak singularity of the pressure, diverging logarithmically with the distance to the tip. 

Panel (a) corresponds to a case without Shuttleworth effect $(c_{0,1}=0)$, for different ratios $\gamma_{LV}/\gamma_0$. With this, we cover a broad range of $\theta_S$ down to very narrow angles with $20^\circ$. The prefactor of the logarithmic pressure singularity is larger for narrow $\theta_S$. The pressure plotted in Fig.~\ref{fig:pressure}(a) is scaled by $G\times\left(\frac{\pi}{\theta_S} -\frac{\theta_S}{\pi} \right)$, which indeed captures the $\theta_S$ dependence of the prefactor of the singularity. We remark that for very narrow angles the logarithmic asymptotic only emerges at distances much below the elastocapillary length $\gamma_0/G$; this illustrates the challenge of accurate numerical resolutions for small $\theta_S$. Panel (b) corresponds to the case with a strong Shuttleworth effect $(c_{0,1}=1)$, for different amounts of substrate prestretch $\lambda_\infty$ (the corresponding $\theta_S$ are in Fig.~\ref{fig:thetavslambda}). Figure~\ref{fig:pressure}(b) again reveals a logarithmic singularity of pressure, with a weak variation of the prefactor with $\lambda_\infty$. 

Interestingly, the Shuttleworth effect allows for a new phenomenon induced by gradients of surface tension. For liquid interfaces, gradients in surface tension arise due to gradients in composition or in temperature -- this is known as the Marangoni effect, and leads to tangential interfacial stress. For the elastic interfaces considered here, the gradients in surface tension are due to gradients of $\lambda$ along the interface. Given this analogy, we refer to this as the \emph{elastic Marangoni effect}.

Figure~\ref{fig:pressure}(c) indeed reveals the emergence of elastic (Cauchy) shear stress $\sigma_{nt}$ along the interface, which we will refer to as elastic Marangoni stress. Somewhat surprisingly, the Marangoni stress is not singular, but converges to a constant value upon approaching the contact line. This elastic Marangoni stress can be positive or negative, depending on the prestretch that is imposed. Without prestretch ($\lambda_\infty=1$), the contact line region will have the largest surface tension, giving a Marangoni stress that is oriented towards the contact line ($\sigma_{nt} < 0$). Conversely, when the imposed $\lambda_\infty$ is large, the contact line region has the smallest surface tension and the Marangoni stress is directed away from the contact line ($\sigma_{nt} < 0$). This is further quantified in Fig.~\ref{fig:marangoni}, where the change of direction of the Marangoni effect is observed to be close to $\lambda_{\rm cl}\approx 1.2$. Indeed, this nearly coincides with the point where $\lambda_{\rm tip} \approx \lambda_\infty$ [cf. Fig.~\ref{fig:thetavslambda}(b)]. So the orientation of the Marangoni stress depends on whether the stretch at the tip is larger or smaller than the stretch imposed at large distance.

\section{Exact nonlinear solutions}\label{sec:theory}

\subsection{Splitting off the singularity}

We will now pursue a fully analytical theory for the numerical observations above. We have seen that the elastic singularity is weak, only logarithmic in the stress, so we first try to split off the singularity. For this, we perform an integration by parts on (\ref{eq:weak}) by writing
\begin{eqnarray}\label{eq:varbis}
\delta \mathcal E 
 &=& \int d^2X \, \left( 
\mathrm{Div}  (\mathbf s \cdot \delta \mathbf x)
- \left( \mathrm{Div} \cdot \mathbf s \right) \cdot \delta \mathbf x\right)
+ \oint dS \, \left( \frac{\partial}{\partial S}\left( \Upsilon \mathbf t  \cdot  \delta \mathbf x\right)  
-  \frac{\partial (\Upsilon \mathbf t)}{\partial S}\cdot \delta \mathbf x \right) 
\end{eqnarray}
The integral over the third term indeed gives point-like contributions
\begin{equation}\label{eq:disc}
\oint dS \,  \frac{\partial}{\partial S}\left( \Upsilon \mathbf t  \cdot  \delta \mathbf x\right)  = 
- \sum_{\mathrm{disc.} \, i} \left[ \Upsilon \mathbf t \right]^+_- \cdot \delta \mathbf x_i,
\end{equation}
where the sum runs over all possible discontinuities along the contour. The term $\mathrm{Div}(\mathbf s \cdot \delta \mathbf x)$ can be brought to the surface using the divergence theorem. For a smooth domain of integration, the divergence theorem holds for any vector field which is in ${\mathcal L}^1$ and whose spatial derivatives are in ${\mathcal L}^1$ \cite{Willem2013}.  This is allowed as long as the corresponding singularity is weaker than $1/|\mathbf X|$. 
\begin{eqnarray}\label{eq:varbis}
\delta \mathcal E 
 &=& 
 -\int d^2X \, \mathrm{Div} \left( \mathbf s \right) \cdot \delta \mathbf x 
+ \oint dS \, \left( \mathbf s \cdot \mathbf N - \frac{\partial (\Upsilon \mathbf t)}{\partial S}\right) \cdot \delta \mathbf x 
- \sum_{\mathrm{disc.} \, i} \left[ \Upsilon \mathbf t \right]^+_- \cdot \delta \mathbf x_i
\nonumber \\
 &=& 
 -\int d^2x \, \mathrm{div}\left( \boldsymbol{\sigma} \right) \cdot \delta \mathbf x 
+ \oint ds \, \left( \boldsymbol{\sigma} \cdot \mathbf n - \frac{\partial (\Upsilon \mathbf t)}{\partial s}\right) \cdot \delta \mathbf x 
- \sum_{\rm disc.\, i} \left[ \Upsilon \mathbf t \right]^+_- \cdot \delta \mathbf x_i,
\end{eqnarray}
where in the last step we transformed the result to the current domain, using the definition of the true stress (or Cauchy stress) according to $\boldsymbol{\sigma}=\mathbf s \cdot \mathbf F^T/ \det(\mathbf F)$.  


The condition of equilibrium, $\delta \mathcal E=\delta \mathcal R$ obtained from (\ref{eq:forcing}),(\ref{eq:varbis}), then splits into bulk, surface and point conditions:

 \begin{eqnarray}
 \mathrm{div}( \boldsymbol{\sigma}  )&=& 0, \quad \mathbf x \in \mathcal D, \label{eq:bulk}
 \\
 \boldsymbol{\sigma} \cdot \mathbf n - \frac{\partial (\Upsilon \mathbf t)}{\partial s} &=& 0, 
 \quad \mathbf x \in \partial \mathcal D,\label{eq:laplacemarangoni}
  \\
 \left[ \Upsilon \mathbf t \right]^+_-  +  \gamma_{LV}  \mathbf t_{LV}   &=& 0, 
 \quad \mathbf x =\mathbf x_{\rm cl}, \label{eq:neumannvariation}
 \end{eqnarray}
where $\mathcal D$ denotes the current domain of the deformed state. Besides the classical elastic stress equilibrium in bulk (\ref{eq:bulk}), the interface condition (\ref{eq:laplacemarangoni}) gives the Marangoni effect where $\sigma_{nt}\equiv \mathbf t \cdot \boldsymbol{\sigma} \cdot \mathbf n$ balances gradients in surface tension $\partial \Upsilon/\partial s$, while the normal component of elastic stress $\sigma_{nn}\equiv \mathbf n \cdot \boldsymbol{\sigma} \cdot \mathbf n$ balances the Laplace pressure. Finally, the Neumann condition appears at the contact line, equation (\ref{eq:neumannvariation}), expressed as a discontinuity of the surface tangents. The only assumption made in the derivation above is that the stress singularity is sufficiently weak for the divergence theorem to be applicable, as is the case for a logarithmic singularity.

\begin{figure}[t]
    \centering
    \includegraphics[width=0.8\textwidth]{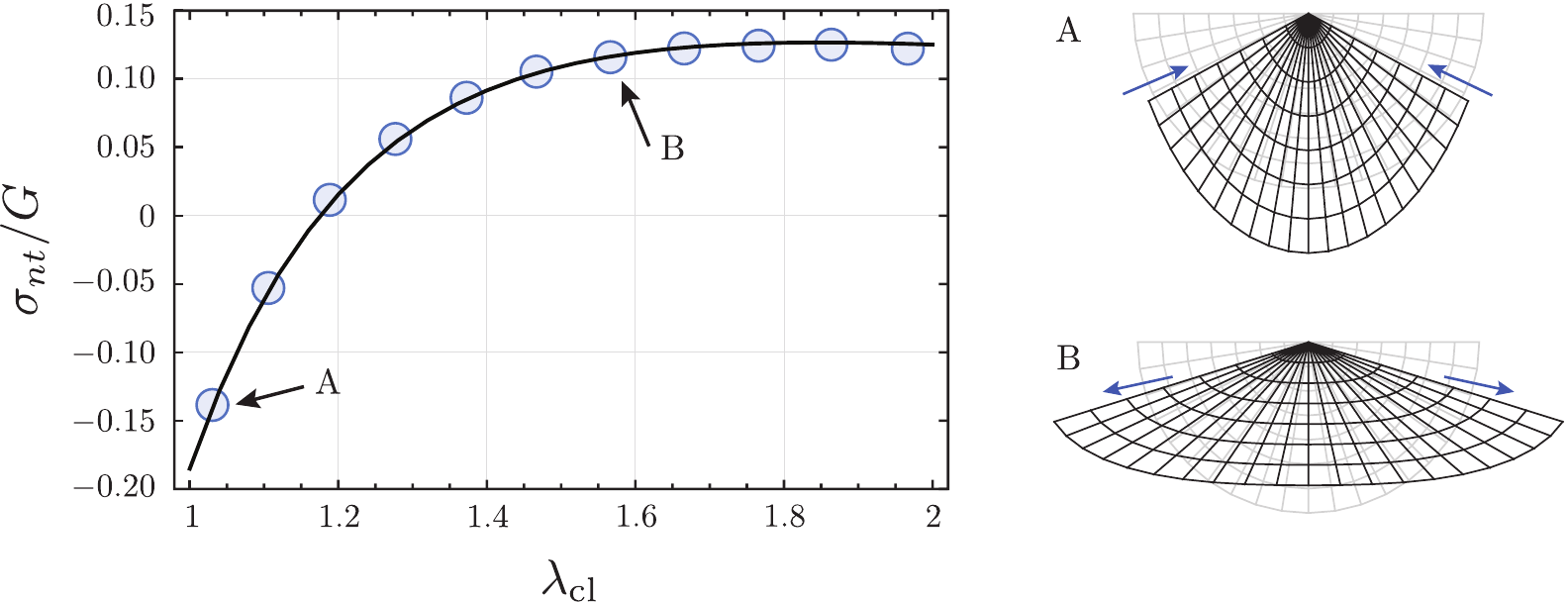}
    \caption{Marangoni stress $\sigma_{nt}/G$ at the contact line, as a function of the stretch at the contact line $\lambda_{\rm cl}$. The Marangoni stress can be positive or negative depending on whether the stretch at the tip is larger or smaller than $\lambda_\infty$. Open circles obtained by FEM, solid line from similarity solutions described in Sec.~\ref{sec:theory}. The grids represent geometry of the ridge and deformation within it for negative (A) and positive (B) Marangoni stresses, as obtained from the similarity solutions. The grey lines denote the undeformed grid, and the arrows indicate the direction of the Marangoni stress. }
           \label{fig:marangoni}
\end{figure}

\subsection{Similarity solutions}


We now analytically establish the nature of the elastic singularity, through an asymptotic analysis near the contact line. For this we express the mapping $\chi(\mathbf X)$ in polar coordinates, $(r,\varphi)$ and $(R,\Phi)$, respectively for the current and reference state. The contact line is located at $r=0$ and $R=0$, and without loss of generality the initially flat free surface is chosen to be along the lines $\Phi=0$ and $\Phi=\pi$. We make use of the fact that the boundary condition (\ref{eq:neumannvariation}) forces the solid into an angle $\theta_S$, which is defined by the property

\begin{equation}\label{eq:defcorner}
\theta_S = \lim_{R\rightarrow 0}\left( \varphi_{\Phi=\pi} - \varphi_{\Phi=0}\right).
\end{equation}
As is common with singularities~\cite{EggersFontelos2015}, we expect the asymptotics to be scale-invariant, so we propose a similarity Ansatz

\begin{eqnarray}
r(R,\Phi)  &=& R^\alpha g_1(\Phi), \nonumber \\
\varphi(R,\Phi)  &=& R^\beta g_2(\Phi).
\end{eqnarray}
Imposing (\ref{eq:defcorner}) one finds that $\beta=0$. A critical feature of soft elastic solids is that these are basically incompressible, i.e. $\mathrm{det}(\mathbf F)=1$. Combined with $\beta=0$, this then dictates $\alpha=1$, which implies that the radial stretch $\lambda_r = dr/dR$ remains finite and is independent of $R$. In the azimuthal direction, incompressibility implies a relation between the functions $g_{1,2}$, which can be accounted for by writing

\begin{eqnarray}
r(R,\Phi)  &=& \frac{R}{\sqrt{f(\Phi)}}, \nonumber \\
\varphi(R,\Phi)  &=& \int_{\Phi_0}^\Phi dU \, f(U),\label{eq:phicorner}
\end{eqnarray}
so that the solid angle follows as $\theta_S =  \int_0^\pi d\Phi \, f(\Phi)$. The deformation gradient tensor of this mapping reads

\begin{equation}
\mathbf F =
\begin{pmatrix}
F_{rR} & F_{r\Phi}\\
F_{\varphi R} & F_{\varphi \Phi} \\
\end{pmatrix}
=
\begin{pmatrix}
\frac{\partial r}{\partial R} & \frac{1}{R}\frac{\partial r}{\partial \Phi}\\
 r\frac{\partial \varphi}{\partial R} &  \frac{r}{R}\frac{\partial \varphi}{\partial \Phi} \\
\end{pmatrix}
=
\begin{pmatrix}
\frac{1}{\sqrt{f}}  
& -\frac{1}{2\sqrt{f}} \frac{f'}{f} \\
0 & \sqrt{f} \\
\end{pmatrix},
\end{equation}
which indeed satisfies $\mathrm{det}(\mathbf F) =1$ for arbitrary $f(\Phi)$. The corresponding Finger tensor reads

\begin{equation}\label{eq:finger}
\mathbf B = \mathbf F \cdot \mathbf F^T =
\begin{pmatrix}
\frac{1}{f}\left(1 + \left(\frac{f'}{2f}\right)^2 \right)  
& - \frac{f'}{2f} \\
- \frac{f'}{2f} & f \\
\end{pmatrix}.
\end{equation}
This defines the most general scale-invariant incompressible map that generates a corner. 

For the special case where $f'=0$, one recovers the classical solution by Singh \& Pipkin~\cite{Singh1965}. However, that solution is shear-free (i.e. $\mathbf F$ and $\mathbf B$ are diagonal) and therefore cannot be universally valid. Here we derive the most general corner solution that satisfies mechanical equilibrium, $\mathrm{div}(\boldsymbol{\sigma})=0$. We focus on a neo-Hookean material defined by (\ref{eq:Wneohookean}), which has a Cauchy stress $\boldsymbol{\sigma} = G \mathbf B - p \mathbf I$, so that (\ref{eq:bulk}) becomes

\begin{equation}\label{eq:nablapressure}
\mathrm{grad}(p) = G \, \mathrm{div}(\mathbf B).
\end{equation}
This implies that $\textrm{div}(\mathbf B)$ must be irrotational, i.e. $\mathrm{curl}\left(\mathrm{div}(\mathbf B) \right)=0$, which here takes the form

\begin{equation}
\frac{\partial }{\partial \varphi}\left(\frac{\partial B_{r\varphi}}{\partial \varphi} + B_{rr}- B_{\varphi \varphi}\right) =0
\quad \Rightarrow \quad 
\frac{\partial B_{r\varphi}}{\partial \varphi} + B_{rr} - B_{\varphi \varphi} =K,
\end{equation}
where $K$ is an integration constant. Inserting (\ref{eq:finger}) and bearing in mind that $\partial/\partial \varphi(\cdots ) =(\cdots)' /f$, we find

\begin{eqnarray}\label{eq:ODE}
- \left(\frac{f'}{2f}\right)' + 1+ \left( \frac{f'}{2f}\right)^2  - f^2 = Kf.
\end{eqnarray}
This is a nonlinear second order ODE for $f(\Phi)$. As boundary conditions we impose the stretch at each of the boundaries, which will subsequently give the shear stress via the connections

\begin{eqnarray}
\lambda_r&=& \frac{dr }{dR} = \frac{1}{\sqrt{f(\Phi)}} \quad \Rightarrow \quad \sigma_{r\varphi} = -\frac{G f'(\Phi)}{2f(\Phi)},
\end{eqnarray}
We note that $\lambda_r$ in the similarity solution is independent of $R$, and can therefore be identified to the stretch at the contact line $\lambda_r = \lambda_{\rm cl}$. The constant $K$ can be adjusted to accommodate the desired $\theta_S$. Explicit solutions will be presented below, and compared directly to FEM simulations. 

Once a solution is found, one can explicitly integrate (\ref{eq:nablapressure}) to obtain the pressure

\begin{equation}\label{eq:plog}
p(r,\varphi) = G K \log r,
\end{equation}
up to an integration constant. This completes the analytical description of incompressible corner solutions in the fully nonlinear regime.

\subsection{Theory compared to FEM}

The similarity solutions derived above capture all FEM results of Sec.~\ref{sec:FEMresults}, in the vicinity of the contact line. First, we consider the stress, which for a neo-Hookean solid is given by $\boldsymbol{\sigma} = G \mathbf B - p \mathbf I$. Our theory explains the FEM result that the normal stress diverges logarithmically, following the singularity of pressure (\ref{eq:plog}), and offers a way to compute the prefactor $K$. Furthermore, the corner solution shows that  $\mathbf B$ as given in (\ref{eq:finger}) remains finite at the contact line. This explains why the Marangoni stress $\sigma_{nt}=\sigma_{r\varphi}$ remains finite at the contact line.

\begin{figure}[t]
    \centering
    \includegraphics[width=1\textwidth]{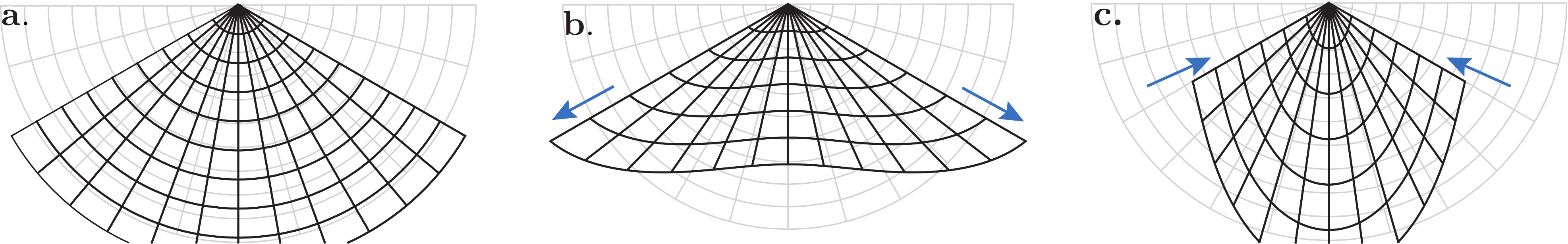}
    \caption{Similarity solutions for symmetric corners obtained from (\ref{eq:phicorner}) and (\ref{eq:ODE}), all with $\theta_S=120^\circ$. (a) Without Shuttleworth effect, the shear stress vanishes at the interface and one recovers the Singh-Pipkin solution~\cite{Singh1965}. (b,c) The Shuttleworth effect induces Marangoni stresses, giving positive (b) or negative (c) elastic shear stress at the interface, the direction indicated by the arrows.}
           \label{fig:grids}
\end{figure}

We now turn to a fully quantitative analysis, by solving (\ref{eq:ODE}) for various boundary conditions. Typical (symmetric) similarity solutions are represented graphically in Fig.~\ref{fig:grids} denoting the Lagrangian grid both in undeformed (grey) and deformed (black) configurations. The three panels each correspond to $\theta_S=120^\circ$, with different amount of stretch imposed on the free surfaces. In Panel (a) we report the solution without shear stress, for which $f'=0$ for all $\varphi$. In this case, (\ref{eq:phicorner}) reduces to the classical solution by Singh \& Pipkin~\cite{Singh1965}, with the constant $f=\theta_S/\pi$. In the context of elastocapillary ridges, the absence of shear corresponds to a substrate without a Shuttleworth effect. This explains why in the absence of a Shuttleworth effect, the stretch at the contact line $\lambda_{\rm cl}$ was found to be independent of $\lambda_\infty$ in our FEM simulations: in a shear-free corner, the stretch takes on a specific value that depends only on the solid angle, as $\lambda_r=\lambda_{\rm cl}=\sqrt{\pi/\theta_S}$. The stretch at the contact line is therefore locally determined by $\theta_S$, irrespective of the conditions imposed at large distance. Furthermore, in this specific case without shear stress, we find an analytical expression for the strength of the pressure singularity, the constant $K$ in (\ref{eq:plog}). Inserting $f=\theta_S/\pi$ in (\ref{eq:ODE}) gives $K=\frac{\pi}{\theta_S}-\frac{\theta_S}{\pi}$. Indeed, this was exactly the scaling used in Fig.~\ref{fig:pressure}(a), necessary to account for the $\theta_S$ dependence. This demonstrates that the corner solutions are fully quantitative and provide the correct asymptotics observed in FEM, valid in the strongly nonlinear regime.

The Shuttleworth effect dramatically changes the physical picture. Now, a variety of surface stretches $\lambda_r$ is possible, as shown in Fig.~\ref{fig:grids}(b,c). Each of these solutions comes with its own value of the elastic Marangoni stress. Figure~\ref{fig:marangoni} illustrates this point, where the prediction of the similarity solutions is shown as a solid line and compared directly to the Marangoni stress in FEM. For the symmetric surface tensions considered in our simulations, the corresponding similarity solution is naturally symmetric and can be found without any adjustable parameters: it follows directly from the surface constitutive relation (\ref{eq:rheology}), which in combination with Neumann's law determines the appropriate combination of $\theta_S$ and $\lambda$. The perfect prediction of the elastic Marangoni stress in Fig.~\ref{fig:marangoni} confirms that the corner solutions indeed offer the correct asymptotic description of the singularity -- also in the presence of the Shuttleworth effect. 

As a conclusive remark, we emphasise again that the observation in PDMS that $\lambda_{\rm cl}$ increases upon varying $\lambda_\infty$  \cite{XuNatComm2017} cannot be explained in a hyperelastic theory without a Shuttleworth effect.

\section{Liquid contact angle, pinning and Eshelby forces}\label{sec:pinning}

\subsection{Hysteresis via a process zone}

So far we have considered an isolated contact line, at some prescribed position $\mathbf X_{\rm cl}$, pulling vertically with perfectly symmetric wetting conditions. In a real wetting problem, however, a droplet will spread dynamically until it reaches its equilibrium liquid angle -- simultaneously the contact line reaches an equilibrium material position $\mathbf X_{\rm cl}$, which is not known a priori but which needs to be found self-consistently. Hence, the full equilibration involves a free exploration of the contact line over the substrate. Technically, such an equilibrium without pinning implies that the change of material coordinate is energetically neutral. 
Naturally, this is the case when the substrate is perfectly homogeneous in its reference state. Indeed, in contrast to the rigid case, there are various examples where well-prepared soft polymeric substrates are basically free from pinning and contact angle hysteresis~\cite{XuNatComm2017,Schulman2018aa,Snoeijer2018,Lhermerout2016aa}.  

Here we take the \emph{opposite} perspective and consider the possibility that the presence of the contact line itself induces heterogeneity in the material -- in its reference state. Even when the originally prepared soft polymeric substrate does not exhibit permanent defects that provide a frozen surface energy landscape, the substrate can develop heterogeneities dynamically,  due to the presence of the contact line. Indeed, a large-stress region builds up at small scale, which can lead to irreversible plastic flow, like in the ``Fracture Process Zone" that forms at a crack tip. Although wetting-induced damaging processes have been evidenced in experiments where a soft gel exhibits fracture by wetting~\cite{bostwick2013capillary}, we focus here on non-damaging plastic deformations in the near-surface region -- so that the bulk reference is not affected. Plasticity typically occurs in situations where there is multistability, where multiple stable configurations coexist, which can lead to a hysteretic response upon contact displacement \cite{Caroli1998}. 
The large strain may indeed provide a configurational plasticity, without damaging the material. When chains between cross-linkers are long enough to produce entanglement, strain may trigger changes of glassy chain conformation. As an alternative mechanism, the contact line may lead to a local strengthening associated with the elongation of polymeric chains, producing a highly dissipative zone when the contact line explores its environment. Below we derive the consequences of a non-damaging, plastic process zone induced by the presence of the contact line. By analogy with fracture mechanics, or with defects in crystalline solids, such a plastic process zone can be described by a defect singularity in the theory of elasticity. The singularity then represents the effect of the plastic process zone on the elastic ``outer" region. We reveal how the strength of such a defect directly relates to contact angle hysteresis.


%
\subsection{Displacing an elastocapillary defect}

\begin{figure}[t]
    \centering
    \includegraphics[width = 0.45\textwidth]{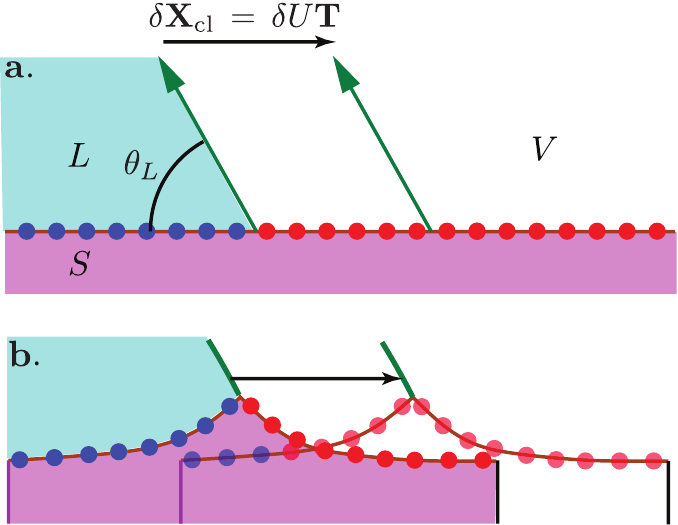}
    \caption{Determining the liquid contact angle $\theta_L$ upon global displacement of the solution. (a) Lagrangian point of view: On the domain of material coordinates the shift is achieved by a change of material point $\delta \mathbf X_{\rm cl} = \delta U \mathbf T$, where the contact line applies. Without pinning, the displacement is energetically neutral, while in the presence of a pinning defect an energy $-\Gamma \delta U$ is dissipated at the contact line. (b) Eulerian point of view: The displacement $\delta U$ leads to a variation of the entire solution as given in (\ref{eq:shift}). At large distance from the contact line, the change of the surface energies reads $(\gamma_{SL}-\gamma_{SV})\lambda_\infty \delta U$.} 
           \label{fig:young}
\end{figure}

The consequence of a defect, representing the effect of a process zone on the outer region, can be computed from the change in energy associated to a global displacement of the solution. This is illustrated in Fig.~\ref{fig:young}, showing such a displacement $\delta \mathbf X_{\rm cl}=\delta U \mathbf T$ on the reference domain (panel a), and on the current domain (panel b). The change in \emph{elastic} energy associated to the displacement of a defect is known as the Eshelby force \cite{Eshelby1975},
\begin{equation}\label{eq:eshelby}
f_{\rm Esh} = - \frac{\partial \mathcal E_{\rm el}}{\partial U} = \mathbf T \cdot \oint dS \, \mathbf \Pi \cdot \mathbf N,
\end{equation}
where the integral encloses the defect and we define the Eshelby's energy-momentum tensor
\begin{equation}\label{eq:pi}
\mathbf \Pi = W \mathbf I - \mathbf F^T \cdot \frac{\partial W}{\partial \mathbf F}.
\end{equation}
The Eshelby force reduces to the J-integral in small deformation (linear) elasticity, where it finds an interpretation as the energy release rate in fracture mechanics~\cite{Rice68}. To derive the \emph{capillary} energy released by moving a defect, it is instructive to follow the derivation of the classical result (\ref{eq:eshelby}), which is based on the application of Noether's theorem in the space of material coordinates \cite{Eshelby1975}. 

On the reference domain, the displacement simply amounts to a translation $\delta \mathbf X_{\rm cl}=\delta U \mathbf T$ of the contact line force, as in Fig.~\ref{fig:young}(a). The corresponding translation on the current domain is sketched in Fig.~\ref{fig:young}(b). The idea of deriving the elastic energy released by displacing a defect is to interpret the translation $\delta U$ as a variation $\delta \mathbf x$, which can be expressed as 

\begin{equation}\label{eq:shift}
\delta \mathbf x = \chi \left( \mathbf X - \delta U \mathbf T \right) - \chi\left( \mathbf X \right)  
=  - \frac{\partial \chi}{\partial \mathbf X} \cdot \mathbf T \delta U = - \delta U \mathbf T \cdot \mathbf F^T.
\end{equation}
The associated change in elastic energy can be computed from this variation, as
\begin{eqnarray}
\delta \mathcal E_{\rm el} &=& 
\int d^2X \,   \delta \mathbf x \cdot \left(\frac{\delta \mathcal E_{\rm el}}{\delta \mathbf x}\right)  =
 - \delta U\, \mathbf T \cdot \int d^2X \,  \left( \mathbf F^T \cdot \frac{\delta \mathcal E_{\rm el}}{\delta \mathbf x} \right)
= - \delta U\, \mathbf T \cdot \int d^2X \, \mathrm{Div}\left(\mathbf \Pi \right).
\end{eqnarray}
Importantly, in the last step one uses that the (reference) substrate is homogeneous everywhere except at the defect \cite{Eshelby1975}. When in the vicinity of the singularity $\mathbf \Pi \sim 1/|\mathbf X-\mathbf X_{\rm cl}|$, the integral is finite and can be expressed as (\ref{eq:eshelby}). When the material is homogeneous everywhere, i.e. no defects,   the Eshelby force uniformly vanishes as a consequence of translational invariance. 

We now follow the same scheme for the capillary energy, upon replacing $W$ by $\lambda \gamma$, and the deformation gradient tensor $\mathbf F$ by its vectorial surface analogue $\mathbf F_s = \partial \mathbf x_s/\partial S$. Subsequently, we define the surface-equivalent of the Eshelby tensor (\ref{eq:pi}), which now is a scalar, and which takes the form:

\begin{equation}
\lambda \gamma - \mathbf F_s^T \cdot  \left(\frac{\partial (\lambda \gamma)}{\partial \mathbf F_s}\right) = 
\lambda \gamma - \lambda \Upsilon = - \lambda^2 \gamma'  \equiv - \mu,
\end{equation}
which is the chemical potential anticipated in (\ref{eq:gammamu}). 
Indeed, the associated change in capillary energy reads

\begin{eqnarray}
\delta \mathcal E_{\rm cap} &=& - 
\delta U \int dS \,  \left( \mathbf F_s^T \cdot \frac{\delta \mathcal E_{\rm cap}}{\delta \mathbf x} \right)
=   \delta U \int dS \, \frac{d\mu}{dS}
= \delta U \left[ \mu \right]^+_-,
\end{eqnarray}
where the integral runs over an infinitesimal domain across the singularity. It is clear that a finite capillary defect-energy appears only when $\mu$ exhibits a discontinuity at the contact line, i.e. $[\mu]_-^+ \neq 0$. 

We thus conclude that the total energy release rate $\Gamma$, liberated upon displacing the elastocapillary defect at the contact line, takes the form

\begin{equation}\label{eq:release}
\Gamma = - \frac{ \partial \mathcal E}{\partial U} = - \left[ \mu \right]^+_- + \mathbf T \cdot \oint dS \, \mathbf \Pi\cdot \mathbf N 
= - \left[ \mu \right]^+_- + f_{\rm Esh}.
\end{equation}
Given that the defect represents a process zone, this indicates a loss of energy $-\Gamma \delta U$, dissipated inside the process zone during the translation. For the special case where there is no pinning defect and the contact line is free to move, the variation of the contact line position should be energetically neutral, so that $\Gamma=0$.

The notion of the (elastic) Eshelby force in wetting was recently proposed in \cite{MasurelPRL2019}, where it was argued that the formation of a ridge would already be sufficient to induce an elastic Eshelby force. However, from the above it is clear that this is not the case when the substrate is perfectly \emph{homogeneous in its reference state}, so that there is a translational invariance of the space of reference coordinates: applying Noether's theorem to this translational invariance \cite{Eshelby1975}, one finds $\partial \mathcal E_{\rm el}/\partial U=0$. This vanishing of the Eshelby force is indeed confirmed by our FEM results and analytical solutions: the stress is only logarithmically singular, so that for an infinitesimal integration volume around the contact line (\ref{eq:eshelby}) gives $f_{\rm Esh}=0$. Therefore, for homogenous substrates, the condition $\Gamma=0$ reduces to $[\mu]_-^+ = 0$. The continuity of $\mu$ across the contact line can be interpreted as an ``equality of chemical potential", necessary for a free exchange of material points across the contact line. This condition of no-pinning was previously derived within the strong restrictions of linear elasticity \cite{Snoeijer2018} -- but it turns out to be valid also when deformations are large. 

For a non-damaging process zone, i.e. the reference state remains intact, we expect the Eshelby force to vanish owing to translational invariance. Nonetheless, a capillary defect $\Gamma$ could still emerge, associated with the interfacial microstate of the polymer.

\subsection{The liquid contact angle}

Up to here we have considered properties of the solid, and did not discuss explicitly the liquid. Yet, the liquid contact angle $\theta_L$ is the prime feature that characterises the wetting of a liquid drop. To complete the theory, we now show how the equilibration determines $\theta_L$ on homogeneous substrates --  and how the maximum strength of a contact line defect can be related to contact angle hysteresis on elastic substrates.

We restrict ourselves to the case of a sufficiently large drop, so that far away from the contact line one encounters a flat substrate (Fig.~\ref{fig:young}). At a large distance from the contact line, the substrate respectively has a solid-liquid energy $\gamma_{SL}(\lambda_\infty)$ and a solid-vapour energy $\gamma_{SV}(\lambda_\infty)$. The usual argument leading to Young's law for the contact angle amounts to the global horizontal displacement~\cite{deGe02}. In the present case the (Eulerian) displacement reads $\lambda_\infty \delta U$, so that the solid capillary energy increases by $(\gamma_{SL}-\gamma_{SV})\lambda_\infty \delta U$, the value of which has to be taken far away from the contact line. This balances the work $-\gamma_{LV}\cos \theta_L  \lambda_\infty \delta U$ performed by the liquid-vapour interface, which together gives Young's law. The situation is modified by the presence of a defect: as described above, such a displacement also involves a dissipation inside the process zone, indicating a loss of energy $-\Gamma \delta U$. By consequence, we find a modification of Young's law 

\begin{eqnarray}
 \lambda_\infty \left(\gamma_{SL}  - \gamma_{SV}\right)_{\lambda_\infty} + \Gamma = - \lambda_\infty \gamma_{LV} \cos \theta_L  
\quad 
\Rightarrow \quad \gamma_{LV} \left(\cos \theta_L - \cos \theta_{Y,\lambda_\infty}\right) =  \lambda_\infty^{-1} \left[\mu \right]_{SL}^{SV}
\label{eq:youngplus}
\end{eqnarray}
where in the second line we anticipate that $f_{\rm Esh}=0$ (owing to the weak logarithmic elastic singularity). For homogeneous substrates $\Gamma =  0$, and we recover Young's law for the liquid contact angle. We remark that, $\theta_Y$ is based on the surface energies corresponding to $\lambda_\infty$. 

The analysis above, in particular (\ref{eq:youngplus}), can be verified by the FEM simulations. In the numerics, we fix a priori the material position $\mathbf X_{\rm cl}$ of the pulling force, so that we effectively work with a pinned contact line. 
For symmetric surface tensions and pulling vertically, this is equivalent to the unpinned case, but we can consider any liquid angle $\theta_L$, by changing the pulling direction $\mathbf t_{LV}=(-\cos \theta_L,\sin \theta_L)$ in (\ref{eq:weak}). We then measure the jump $[\mu]_-^+=[\mu]_{SL}^{SV}$ across the contact line obtained for the corresponding solution, as a function of $\theta_L$. We consider two cases: (i) symmetric surface energies $\gamma_{SL} = \gamma_{SV}$ (so that $\theta_Y=90^\circ$), and (ii) asymmetric surface energies $\gamma_{SL} \neq \gamma_{SV}$ (here with $\theta_{Y}=113.6^\circ$). 

\begin{figure}[t]
    \centering
    \includegraphics[width=0.85\textwidth]{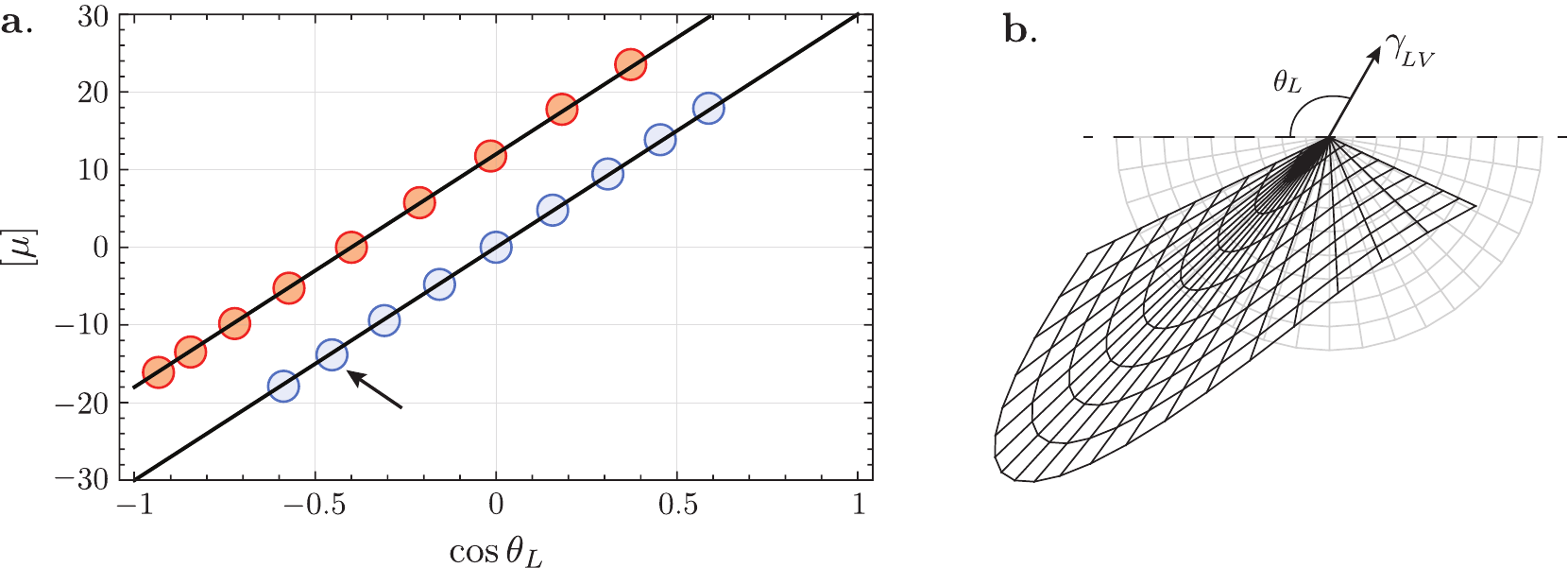}
    \caption{(a) The strength of the surface defect, quantified by the discontinuity of chemical potential $[\mu]_-^+$, plotted versus the liquid contact angle $\theta_L$.  Solid lines are theory of (\ref{eq:youngplus}), open symbols from FEM with Shuttleworth effect ($c_0=c_1=1$). Blue data: symmetric surface energies $\gamma_{0,SV}=\gamma_{0,SL}=\gamma_{LV}$, so that the equilibrium angle $\theta_Y=90^\circ$. Red data: asymmetric surface energies $\gamma_{0,SV}=4/5\gamma_{LV}$ and $\gamma_{0,SL}=6/5\gamma_{LV}$, so that based on $\lambda_\infty =1$ we find $\cos \theta_Y = -2/5$. The numerics confirm that  $[\mu]_-^+$ provides the pinning force; when pulling at $\theta_L=\theta_Y$ there is no pinning and $[\mu]_-^+=0$. (b) The grid plot represents the asymmetric ridge for symmetric surface energies, resulting from a contact angle $\theta_L>\theta_Y$. This ridge corresponds to the data point marked by an arrow in the main panel.}
           \label{fig:pinning}
\end{figure}

The result is presented in Fig.~\ref{fig:pinning}(a). It is clear that both cases, symmetric and asymmetric, are in perfect agreement with (\ref{eq:youngplus}) with $\Gamma = - [\mu]_{SL}^{SV}$. This implies that $f_{\rm Esh}=0$, consistent with the weak logarithmic singularity. Hence, $\theta_L$ can be different from its equilibrium value $\theta_Y$ by the presence of a non-damaging process zone, represented by a capillary defect. In that case, interfacial plasticity could be associated with a contact angle hysteresis. A typical asymmetric similarity solution is shown via the grid representation in Fig.~\ref{fig:pinning}(b), for which there is a jump in stretch at  across the contact line. We remark that in all cases, Neumann's law was still observed to be valid, irrespective of the defect.

\section{Discussion}

In summary, we have explored analytically and numerically the macroscopic theory for elastocapillary ridges, based on the minimisation of a bulk elastic free energy and a surface capillary free energy. This for the first time offers a fully self-consistent description of ``Soft Wetting", including the possibility that capillarity depends on strain (Shuttleworth effect), large elastic deformation, and pinning. In this macroscopic theory there is a perfect separation of scales between elastocapillary length $\gamma/G$ and the molecular scale $a$, since effectively $a\rightarrow 0$ in the continuum. This limit is relevant for typical experiments, and it is of theoretical importance in order to reveal the nature of the ridge-singularity as predicted from large deformation elasticity. We now discuss these new theoretical results in comparison to recent literure on the Shuttleworth effect.

\subsection{Theory}

\textbf{First boundary condition.} 
In this macroscopic description, it was found that the stress singularity associated with the contact line ridge is weak (i.e. logarithmic) and therefore integrable, under all conditions that were considered. Hence, the singularity does not behave analogously to an elastic disclination defect and no qualitative difference emerges when the substrate is globally stretched. As a consequence, in this limit where $\gamma/Ga \to \infty$, the Neumann tension balance at the contact line is strictly valid. In the scheme of energy minimisation, Neumann's law emerges as auxiliary condition (\ref{eq:neumannvariation}), and as such  serves as a \emph{first boundary condition} at the contact line. We have no explanation why previous continuum simulations suggested a deviation from Neumann's law \cite{Wu2018,MasurelPRL2019}. We emphasise, however, that the present numerics are based on an adaptive method, which was necessary to fully resolve the elastic singularity, and that we extensively verified that the results are fully converged. Furthermore we derived new analytical solutions of nonlinear elasticity that describe the singularity -- these are indeed perfectly recovered by the numerics. 

How can one understand the deviation to Neumann's law observed in molecular dynamics simulations of wetting on cross linked polymer networks \cite{liang2018surface}? This deviation finds its origin in the lack of scale separation between $\gamma/G$ and the molecular scale $a$, which is inevitable in molecular simulations -- the scale $a$ there enters as a molecular cutoff of the continuum and also gives a finite width of the interface. As argued in \cite{SJHD2017,AS2020}, the elastic contributions near the contact line can be computed by integrating the elastic stress over a small but finite region -- in molecular simulations, the smallest possible size for this region would be $a$. In the present work we have demonstrated that the stress singularity is always logarithmic, $\sigma  \sim G \log(r/\frac{\gamma}{G})$. Hence, the integral over stress gives an elastic contribution 

\begin{equation}\label{eq:MD}
\int_{-a}^a dr \, \sigma \sim Ga \log(aG/\gamma),
\end{equation}
which needs to be compared to the surface tensions. In molecular simulations, where typically $\gamma/Ga$ is of order 1 to 100, a measurable elastic correction to Neumann's law indeed appears. We refer to~\cite{AS2020} for a quantitative test of (\ref{eq:MD}), as the elastic correction to Neumann's law. However, in typical experiments performed with polymeric gel, where the $\gamma/G$ is well above the micron scale, the scale-separation correction would be $10^{-4}$, and one approaches the macroscopic continuum limit. In such experiments, one safely concludes that Neumann's law holds.\\

\textbf{Second boundary condition versus pinning.} While much theoretical work focussed on the validity (or not) of  Neumann's law, very little attention was given to the implications of contact line pinning \cite{AS2020}. In many experiments on soft polymer networks contact line pinning is virtually absent, as quantified by a very small hysteresis \cite{XuNatComm2017,Schulman2018aa,Snoeijer2018,Lhermerout2016aa}. This implies that the contact line can freely move, exchanging the substrate's material point touching the liquid-vapour interface without any energetic cost. Here we demonstrated that such a free motion occurs only under a very specific condition. Namely the chemical potential defined by $\mu = \lambda^2 d\gamma/d\lambda$ must be continuous across the contact line. This is the \emph{second boundary condition} that needs to be imposed when there is no contact line pinning. Such a condition was previously derived under the restrictive assumption of linear elasticity~\cite{Snoeijer2018} -- here we demonstrate this to be valid also at large deformation, and explored its consequences in numerical simulations. In particular, we have confirmed numerically that Young's law is only recovered when the second boundary condition, $\mu_{SV}=\mu_{SL}$, is satisfied at the contact line. For asymmetric surface energies, the second boundary condition in general implies a jump in stretch across the contact line, so that in the presence of a Shuttleworth effect one generically expects large deformations. 

The possibility of pinning is interesting in itself. Depending on the material strength, large deformations might lead to fracture, as observed in \cite{bostwick2013capillary}, or local  plasticity. We demonstrated how such a local ``process zone" can be accounted for by introducing a defect in the elastocapillary continuum theory. With the defect, one can accommodate a range of angles $\theta_L$ by adjusting the strength of the defect at the contact line. In our simulations we only encountered a weak logarithmic singularity of elastic stress, which implies that the strength of the defect received no contribution from elasticity (i.e. the elastic Eshelby force vanishes). The defect strength, in fact, was found to be equal to the discontinuity in chemical potential at the contact line, i.e. $\Gamma = - [\mu]_-^+$, giving rise to a modified Young's law (\ref{eq:youngplus}). In practice, one would expect the defect to exhibit a ``toughness", just like in fracture, which is the maximum value that can be sustained before depinning occurs. Given (\ref{eq:youngplus}), we immediately infer that this implies a contact angle hysteresis, with advancing and receding angles $\cos \theta_r - \cos \theta_a = 2|[\mu]_{-}^+|_{\rm max}$. Future theoretical work should be dedicated to a more detailed description of the interior of the process zone.

\subsection{Experiments and outlook}

Experiments that probe the strain-dependence of surface tension have so far been based on wetting experiments, with a key role to the contact angles of the solid and of the liquid. Having established the elastocapillary continuum framework for soft wetting, for the first time consistently accounting for large deformations and the Shuttleworth effect, we can now critically assess the experimental situation. 

Different series of experiments have been performed with stretched PDMS gels, for static \cite{XuNatComm2017,xu2018} and dynamical wetting \cite{Snoeijer2018}. They consistently show a change of the solid angle $\theta_S$ under stretching. Similarly, the solid angle was found to change in dynamical experiments on PVS \cite{Gorcum2020}. Till date, these experimental observations have not received any other explanation than via a surface tension that depends on the strain (or, on the history of strain). Hence, they offer a convincing case for a nontrivial surface constitutive relation in soft polymer networks, at least for two different systems. 

Another direct piece of evidence for the Shuttleworth effect is that experiments in \cite{XuNatComm2017} reveal an increase of stretch at the contact line upon a global stretching of the substrate. This information was previously not used to interpret results in the context of a Shuttleworth effect. However, our numerical and analytical results show that such a variation of the stretch at the contact line can only occur in the presence of elastic Marangoni stresses, induced by a Shuttleworth effect -- if surface tension were constant, the stretch at the contact line would take on a constant value.

This evokes an important question that remains to be resolved: What is the microscopic origin of the coupling between surface energy and strain? The polymer is expected to be liquid-like at small scale, where surface tension is exerted: What can produce the coupling between the microscopic scale and the deformation of the network of reticulation (or entanglement) points? A possible scenario is that the coupling emerges from a superficial layer where the mechanical and structural properties are different from bulk~\cite{AS16,Schulman2018aa}. Related to this open question is the experimental observation that, in contrast to solid angle $\theta_S$, the liquid contact angle $\theta_L$ turns out \emph{not} to depend on stretching \cite{Schulman2018aa} -- a property that was confirmed for 6 different liquid-substrate systems in \cite{Schulman2018aa}, and which also holds for PDMS \cite{XuNatComm2017,Snoeijer2018}, for substrates stretched up to 100\%. This is surprising, since Young's law for the liquid angle should be valid for sufficiently large drops, but with surface energies $\gamma_{SV}-\gamma_{SL}$ based on the externally imposed stretch $\lambda_\infty$~\cite{Schulman2018aa}. This interpretation of Young's law is confirmed in Sec.~\ref{sec:pinning}, in an analysis where the large deformation elasticity and the Shuttleworth effect are explicitly  accounted for. The implication of the experimental invariance of $\theta_L$ (within an experimental resolution of $\pm 1^\circ$) is that, for all imposed $\lambda_\infty$, the strain dependence $d\gamma/d\lambda$ must be nearly the same on both sides of the contact line. While there is no understanding of the microscopic/mesoscopic origin of the strain dependent surface energy, there is {\it a fortiori} no real understanding of this property, observed to be valid for many different pairs of liquid and reticulated polymers. 

Another assessment of the Shuttleworth effect makes use of an elastic Wilhelmy plate, where a polymeric wire is partially immersed in a liquid reservoir -- allowing one to measure the stretch discontinuity across the contact line. In \cite{ChenDanielsSM2019}, it was found that the strain remains very small and no discontinuity was observed -- implying once more that $d\gamma/d\lambda$ is equal on both sides. In the initial experiment in~\cite{Marchand2012c}, conversely, a strong discontinuity of strain was observed at the contact line, implying a jump in $d\gamma/d\lambda$. Given that strains remain very small in these experiments, we can assume that the measured strain reflects the actual strains close to the contact line. Therefore, one can interpret these experiments using the no-pinning condition of Sec.~\ref{sec:pinning}, i.e. $[\mu]_-^+=0$, which at small strains implies the continuity of $d\gamma/d\lambda$ across the contact line -- in perfect agreement with the observation in \cite{ChenDanielsSM2019}. It was argued in~\cite{ChenDanielsSM2019} that discontinuous strains could be an artefact due to swelling. As an alternative interpretation, we note that in~\cite{Marchand2012c} a strong contact angle hysteresis was observed, which in the Shuttleworth-interpretation would also be consistent with a breakdown of the no-pinning condition $[\mu]_-^+=0$. 

In conclusion, this research opens the promising perspective of identifying different conditions or different preparation protocols to get, or not, polymer networks with intricate surface properties. The main open question is to understand the microscopic origin of the Shuttleworth effect, which in the present understanding is unambiguously confirmed for at least two different systems. We emphasise that mechanically, none of the experimental observations are in contradiction with the presence of a Shutttleworth effect, in particular since $\theta_L$ and the elastic Wilhelmy plate only probe the \emph{difference} of strain-dependence on either sides of the contact line. By contrast, the independent measurements of \emph{both} the solid angle and the stretch at the contact line \cite{XuNatComm2017,xu2018} cannot be explained by a hyperelastic theory without explicitly accounting for a strong Shuttleworth effect. Future experiments on a broad class of soft materials should therefore simultaneously explore both contact angles $\theta_L$ and $\theta_S$, as well as the strains near the contact line. Combined with the fully nonlinear numerics as presented here, this will offer a systematic quantification of the capillarity of soft solids. A next step is to extend the numerical method to a ridge travelling at constant velocity, including the substrate's bulk viscoelasticity, and possibly history-dependent surface rheology \cite{Gorcum2018}. 

\emph{Acknowledgments.}~
We thank J. Eggers, M. van Gorcum, T. Salez and R. Style for discussions. We acknowledge financial support from ERC (European Research Council) Consolidator Grant number 616918,  (to A.P. and S.K.), from ANR (French National Agency for Research) grant SMART (to B.A.), from NWO through VICI Grant No. 680-47-632 (to J.H.S.) and an Industrial Partnership Program (a joint research program of Canon Production Printing, Eindhoven University of Technology,
University of Twente, and NWO (to E.H.B.).


\begin{thebibliography}{54}
\expandafter\ifx\csname natexlab\endcsname\relax\def\natexlab#1{#1}\fi
\expandafter\ifx\csname bibnamefont\endcsname\relax
  \def\bibnamefont#1{#1}\fi
\expandafter\ifx\csname bibfnamefont\endcsname\relax
  \def\bibfnamefont#1{#1}\fi
\expandafter\ifx\csname citenamefont\endcsname\relax
  \def\citenamefont#1{#1}\fi
\expandafter\ifx\csname url\endcsname\relax
  \def\url#1{\texttt{#1}}\fi
\expandafter\ifx\csname urlprefix\endcsname\relax\def\urlprefix{URL }\fi
\providecommand{\bibinfo}[2]{#2}
\providecommand{\eprint}[2][]{\url{#2}}

\bibitem[{\citenamefont{Li et~al.}(2017)\citenamefont{Li, Celiz, Yang, Yang,
  Wamala, Whyte, Seo, Vasilyev, Vlassak, Suo et~al.}}]{Li378}
\bibinfo{author}{\bibfnamefont{J.}~\bibnamefont{Li}},
  \bibinfo{author}{\bibfnamefont{A.~D.} \bibnamefont{Celiz}},
  \bibinfo{author}{\bibfnamefont{J.}~\bibnamefont{Yang}},
  \bibinfo{author}{\bibfnamefont{Q.}~\bibnamefont{Yang}},
  \bibinfo{author}{\bibfnamefont{I.}~\bibnamefont{Wamala}},
  \bibinfo{author}{\bibfnamefont{W.}~\bibnamefont{Whyte}},
  \bibinfo{author}{\bibfnamefont{B.~R.} \bibnamefont{Seo}},
  \bibinfo{author}{\bibfnamefont{N.~V.} \bibnamefont{Vasilyev}},
  \bibinfo{author}{\bibfnamefont{J.~J.} \bibnamefont{Vlassak}},
  \bibinfo{author}{\bibfnamefont{Z.}~\bibnamefont{Suo}}, \bibnamefont{et~al.},
  \bibinfo{journal}{Science} \textbf{\bibinfo{volume}{357}},
  \bibinfo{pages}{378} (\bibinfo{year}{2017}), ISSN \bibinfo{issn}{0036-8075},
  \eprint{https://science.sciencemag.org/content/357/6349/378.full.pdf},
  \urlprefix\url{https://science.sciencemag.org/content/357/6349/378}.

\bibitem[{\citenamefont{Shirtcliffe et~al.}(2011)\citenamefont{Shirtcliffe,
  McHale, and I.~Newton}}]{Newton2011}
\bibinfo{author}{\bibfnamefont{N.~J.} \bibnamefont{Shirtcliffe}},
  \bibinfo{author}{\bibfnamefont{G.}~\bibnamefont{McHale}}, \bibnamefont{and}
  \bibinfo{author}{\bibfnamefont{M.}~\bibnamefont{I.~Newton}},
  \bibinfo{journal}{Journal of Polymer Science Part B: Polymer Physics}
  \textbf{\bibinfo{volume}{49}}, \bibinfo{pages}{1203} (\bibinfo{year}{2011}),
  \eprint{https://onlinelibrary.wiley.com/doi/pdf/10.1002/polb.22286},
  \urlprefix\url{https://onlinelibrary.wiley.com/doi/abs/10.1002/polb.22286}.

\bibitem[{\citenamefont{Rogers et~al.}(2010)\citenamefont{Rogers, Someya, and
  Huang}}]{Rogers1603}
\bibinfo{author}{\bibfnamefont{J.~A.} \bibnamefont{Rogers}},
  \bibinfo{author}{\bibfnamefont{T.}~\bibnamefont{Someya}}, \bibnamefont{and}
  \bibinfo{author}{\bibfnamefont{Y.}~\bibnamefont{Huang}},
  \bibinfo{journal}{Science} \textbf{\bibinfo{volume}{327}},
  \bibinfo{pages}{1603} (\bibinfo{year}{2010}), ISSN \bibinfo{issn}{0036-8075},
  \eprint{https://science.sciencemag.org/content/327/5973/1603.full.pdf},
  \urlprefix\url{https://science.sciencemag.org/content/327/5973/1603}.

\bibitem[{\citenamefont{Discher et~al.}(2005)\citenamefont{Discher, Janmey, and
  Wang}}]{Discher2005aa}
\bibinfo{author}{\bibfnamefont{D.}~\bibnamefont{Discher}},
  \bibinfo{author}{\bibfnamefont{P.}~\bibnamefont{Janmey}}, \bibnamefont{and}
  \bibinfo{author}{\bibfnamefont{Y.}~\bibnamefont{Wang}},
  \bibinfo{journal}{Science} \textbf{\bibinfo{volume}{310}},
  \bibinfo{pages}{1139} (\bibinfo{year}{2005}).

\bibitem[{\citenamefont{Douezan et~al.}(2012)\citenamefont{Douezan, Dumond, and
  Brochard-Wyart}}]{Douezan2012aa}
\bibinfo{author}{\bibfnamefont{S.}~\bibnamefont{Douezan}},
  \bibinfo{author}{\bibfnamefont{J.}~\bibnamefont{Dumond}}, \bibnamefont{and}
  \bibinfo{author}{\bibfnamefont{F.}~\bibnamefont{Brochard-Wyart}},
  \bibinfo{journal}{Soft Matter} \textbf{\bibinfo{volume}{8}},
  \bibinfo{pages}{4578} (\bibinfo{year}{2012}).

\bibitem[{\citenamefont{King et~al.}(2014)\citenamefont{King, Bartlett, Gilman,
  Irschick, and Crosby}}]{King2014}
\bibinfo{author}{\bibfnamefont{D.~R.} \bibnamefont{King}},
  \bibinfo{author}{\bibfnamefont{M.~D.} \bibnamefont{Bartlett}},
  \bibinfo{author}{\bibfnamefont{C.~A.} \bibnamefont{Gilman}},
  \bibinfo{author}{\bibfnamefont{D.~J.} \bibnamefont{Irschick}},
  \bibnamefont{and} \bibinfo{author}{\bibfnamefont{A.~J.}
  \bibnamefont{Crosby}}, \bibinfo{journal}{Advanced Materials}
  \textbf{\bibinfo{volume}{26}}, \bibinfo{pages}{4345} (\bibinfo{year}{2014}),
  \eprint{https://onlinelibrary.wiley.com/doi/pdf/10.1002/adma.201306259},
  \urlprefix\url{https://onlinelibrary.wiley.com/doi/abs/10.1002/adma.201306259}.

\bibitem[{\citenamefont{Zou et~al.}(2018)\citenamefont{Zou, Zhu, Li, Lei,
  Zhang, and Xiao}}]{Zoueaaq0508}
\bibinfo{author}{\bibfnamefont{Z.}~\bibnamefont{Zou}},
  \bibinfo{author}{\bibfnamefont{C.}~\bibnamefont{Zhu}},
  \bibinfo{author}{\bibfnamefont{Y.}~\bibnamefont{Li}},
  \bibinfo{author}{\bibfnamefont{X.}~\bibnamefont{Lei}},
  \bibinfo{author}{\bibfnamefont{W.}~\bibnamefont{Zhang}}, \bibnamefont{and}
  \bibinfo{author}{\bibfnamefont{J.}~\bibnamefont{Xiao}},
  \bibinfo{journal}{Science Advances} \textbf{\bibinfo{volume}{4}}
  (\bibinfo{year}{2018}),
  \eprint{https://advances.sciencemag.org/content/4/2/eaaq0508.full.pdf},
  \urlprefix\url{https://advances.sciencemag.org/content/4/2/eaaq0508}.

\bibitem[{\citenamefont{Binder and Kob}(2011)}]{binder2011glassy}
\bibinfo{author}{\bibfnamefont{K.}~\bibnamefont{Binder}} \bibnamefont{and}
  \bibinfo{author}{\bibfnamefont{W.}~\bibnamefont{Kob}},
  \emph{\bibinfo{title}{Glassy materials and disordered solids: An introduction
  to their statistical mechanics}} (\bibinfo{publisher}{World scientific},
  \bibinfo{year}{2011}).

\bibitem[{\citenamefont{De~Gennes}(1979)}]{de1979scaling}
\bibinfo{author}{\bibfnamefont{P.-G.} \bibnamefont{De~Gennes}},
  \emph{\bibinfo{title}{Scaling concepts in polymer physics}}
  (\bibinfo{publisher}{Cornell university press}, \bibinfo{year}{1979}).

\bibitem[{\citenamefont{Doi and Edwards}(1988)}]{doi1988theory}
\bibinfo{author}{\bibfnamefont{M.}~\bibnamefont{Doi}} \bibnamefont{and}
  \bibinfo{author}{\bibfnamefont{S.~F.} \bibnamefont{Edwards}},
  \emph{\bibinfo{title}{The theory of polymer dynamics}},
  vol.~\bibinfo{volume}{73} (\bibinfo{publisher}{oxford university press},
  \bibinfo{year}{1988}).

\bibitem[{\citenamefont{Rubinstein et~al.}(2003)\citenamefont{Rubinstein, Colby
  et~al.}}]{rubinstein2003polymer}
\bibinfo{author}{\bibfnamefont{M.}~\bibnamefont{Rubinstein}},
  \bibinfo{author}{\bibfnamefont{R.~H.} \bibnamefont{Colby}},
  \bibnamefont{et~al.}, \emph{\bibinfo{title}{Polymer physics}},
  vol.~\bibinfo{volume}{23} (\bibinfo{publisher}{Oxford university press New
  York}, \bibinfo{year}{2003}).

\bibitem[{\citenamefont{Shuttleworth}(1950)}]{Shut50}
\bibinfo{author}{\bibfnamefont{R.}~\bibnamefont{Shuttleworth}},
  \bibinfo{journal}{Proc. Phys. Soc., London Sect. A}
  \textbf{\bibinfo{volume}{63}}, \bibinfo{pages}{444} (\bibinfo{year}{1950}).

\bibitem[{\citenamefont{Andreotti and Snoeijer}({2016})}]{AS16}
\bibinfo{author}{\bibfnamefont{B.}~\bibnamefont{Andreotti}} \bibnamefont{and}
  \bibinfo{author}{\bibfnamefont{J.~H.} \bibnamefont{Snoeijer}},
  \bibinfo{journal}{{EPL}} \textbf{\bibinfo{volume}{{109}}},
  \bibinfo{pages}{{66001}} (\bibinfo{year}{{2016}}).

\bibitem[{\citenamefont{Style et~al.}(2017)\citenamefont{Style, Jagota, Hui,
  and Dufresne}}]{SJHD2017}
\bibinfo{author}{\bibfnamefont{R.~W.} \bibnamefont{Style}},
  \bibinfo{author}{\bibfnamefont{A.}~\bibnamefont{Jagota}},
  \bibinfo{author}{\bibfnamefont{C.-Y.} \bibnamefont{Hui}}, \bibnamefont{and}
  \bibinfo{author}{\bibfnamefont{E.~R.} \bibnamefont{Dufresne}},
  \bibinfo{journal}{Annual Review of Condensed Matter Physics}
  \textbf{\bibinfo{volume}{8}}, \bibinfo{pages}{99} (\bibinfo{year}{2017}).

\bibitem[{\citenamefont{Andreotti and Snoeijer}(2020)}]{AS2020}
\bibinfo{author}{\bibfnamefont{B.}~\bibnamefont{Andreotti}} \bibnamefont{and}
  \bibinfo{author}{\bibfnamefont{J.~H.} \bibnamefont{Snoeijer}},
  \bibinfo{journal}{Annual Review of Fluid Mechanics}
  \textbf{\bibinfo{volume}{52}}, \bibinfo{pages}{285} (\bibinfo{year}{2020}).

\bibitem[{\citenamefont{Marchand
  et~al.}(2012{\natexlab{a}})\citenamefont{Marchand, Das, Snoeijer, and
  Andreotti}}]{Marchand2012c}
\bibinfo{author}{\bibfnamefont{A.}~\bibnamefont{Marchand}},
  \bibinfo{author}{\bibfnamefont{S.}~\bibnamefont{Das}},
  \bibinfo{author}{\bibfnamefont{J.~H.} \bibnamefont{Snoeijer}},
  \bibnamefont{and}
  \bibinfo{author}{\bibfnamefont{B.}~\bibnamefont{Andreotti}},
  \bibinfo{journal}{Phys. Rev. Lett.} \textbf{\bibinfo{volume}{108}},
  \bibinfo{pages}{094301} (\bibinfo{year}{2012}{\natexlab{a}}).

\bibitem[{\citenamefont{Bostwick et~al.}(2014)\citenamefont{Bostwick, Shearer,
  and Daniels}}]{Bostwick:2014aa}
\bibinfo{author}{\bibfnamefont{J.}~\bibnamefont{Bostwick}},
  \bibinfo{author}{\bibfnamefont{M.}~\bibnamefont{Shearer}}, \bibnamefont{and}
  \bibinfo{author}{\bibfnamefont{K.}~\bibnamefont{Daniels}},
  \bibinfo{journal}{Soft Matter} \textbf{\bibinfo{volume}{10}},
  \bibinfo{pages}{7361} (\bibinfo{year}{2014}).

\bibitem[{\citenamefont{Xu et~al.}({2017})\citenamefont{Xu, Jensen,
  Boltyanskiy, Sarfat, Style, and Dufresne}}]{XuNatComm2017}
\bibinfo{author}{\bibfnamefont{Q.}~\bibnamefont{Xu}},
  \bibinfo{author}{\bibfnamefont{K.}~\bibnamefont{Jensen}},
  \bibinfo{author}{\bibfnamefont{R.}~\bibnamefont{Boltyanskiy}},
  \bibinfo{author}{\bibfnamefont{R.}~\bibnamefont{Sarfat}},
  \bibinfo{author}{\bibfnamefont{R.~W.} \bibnamefont{Style}}, \bibnamefont{and}
  \bibinfo{author}{\bibfnamefont{E.~R.} \bibnamefont{Dufresne}},
  \bibinfo{journal}{{Nature Comm.}} \textbf{\bibinfo{volume}{{8}}},
  \bibinfo{pages}{{555}} (\bibinfo{year}{{2017}}).

\bibitem[{\citenamefont{Xu et~al.}(2018)\citenamefont{Xu, Style, and
  Dufresne}}]{xu2018}
\bibinfo{author}{\bibfnamefont{Q.}~\bibnamefont{Xu}},
  \bibinfo{author}{\bibfnamefont{R.~W.} \bibnamefont{Style}}, \bibnamefont{and}
  \bibinfo{author}{\bibfnamefont{E.~R.} \bibnamefont{Dufresne}},
  \bibinfo{journal}{Soft Matter} \textbf{\bibinfo{volume}{14}},
  \bibinfo{pages}{916} (\bibinfo{year}{2018}).

\bibitem[{\citenamefont{Schulman et~al.}(2018)\citenamefont{Schulman, Trejo,
  Salez, Rapha{\"e}l, and Dalnoki-Veress}}]{Schulman2018aa}
\bibinfo{author}{\bibfnamefont{R.~D.} \bibnamefont{Schulman}},
  \bibinfo{author}{\bibfnamefont{M.}~\bibnamefont{Trejo}},
  \bibinfo{author}{\bibfnamefont{T.}~\bibnamefont{Salez}},
  \bibinfo{author}{\bibfnamefont{E.}~\bibnamefont{Rapha{\"e}l}},
  \bibnamefont{and}
  \bibinfo{author}{\bibfnamefont{K.}~\bibnamefont{Dalnoki-Veress}},
  \bibinfo{journal}{Nature Communications} \textbf{\bibinfo{volume}{9}},
  \bibinfo{pages}{982} (\bibinfo{year}{2018}),
  \urlprefix\url{https://doi.org/10.1038/s41467-018-03346-1}.

\bibitem[{\citenamefont{Snoeijer et~al.}(2018)\citenamefont{Snoeijer, Rolley,
  and Andreotti}}]{Snoeijer2018}
\bibinfo{author}{\bibfnamefont{J.~H.} \bibnamefont{Snoeijer}},
  \bibinfo{author}{\bibfnamefont{E.}~\bibnamefont{Rolley}}, \bibnamefont{and}
  \bibinfo{author}{\bibfnamefont{B.}~\bibnamefont{Andreotti}},
  \bibinfo{journal}{Physical Review Letters} \textbf{\bibinfo{volume}{121}}
  (\bibinfo{year}{2018}).

\bibitem[{\citenamefont{Liang et~al.}(2018)\citenamefont{Liang, Cao, Wang, and
  Dobrynin}}]{liang2018surface}
\bibinfo{author}{\bibfnamefont{H.}~\bibnamefont{Liang}},
  \bibinfo{author}{\bibfnamefont{Z.}~\bibnamefont{Cao}},
  \bibinfo{author}{\bibfnamefont{Z.}~\bibnamefont{Wang}}, \bibnamefont{and}
  \bibinfo{author}{\bibfnamefont{A.~V.} \bibnamefont{Dobrynin}},
  \bibinfo{journal}{Langmuir} \textbf{\bibinfo{volume}{34}},
  \bibinfo{pages}{7497} (\bibinfo{year}{2018}).

\bibitem[{\citenamefont{Wu et~al.}(2018)\citenamefont{Wu, Liu, Jagota, and
  Hui}}]{Wu2018}
\bibinfo{author}{\bibfnamefont{H.}~\bibnamefont{Wu}},
  \bibinfo{author}{\bibfnamefont{Z.}~\bibnamefont{Liu}},
  \bibinfo{author}{\bibfnamefont{A.}~\bibnamefont{Jagota}}, \bibnamefont{and}
  \bibinfo{author}{\bibfnamefont{C.-Y.} \bibnamefont{Hui}},
  \bibinfo{journal}{Soft matter} \textbf{\bibinfo{volume}{14}},
  \bibinfo{pages}{1847} (\bibinfo{year}{2018}).

\bibitem[{\citenamefont{Masurel et~al.}(2019)\citenamefont{Masurel, Roch\'e,
  Limat, Ionescu, and Dervaux}}]{MasurelPRL2019}
\bibinfo{author}{\bibfnamefont{R.}~\bibnamefont{Masurel}},
  \bibinfo{author}{\bibfnamefont{M.}~\bibnamefont{Roch\'e}},
  \bibinfo{author}{\bibfnamefont{L.}~\bibnamefont{Limat}},
  \bibinfo{author}{\bibfnamefont{I.}~\bibnamefont{Ionescu}}, \bibnamefont{and}
  \bibinfo{author}{\bibfnamefont{J.}~\bibnamefont{Dervaux}},
  \bibinfo{journal}{Phys. Rev. Lett.} \textbf{\bibinfo{volume}{122}},
  \bibinfo{pages}{248004} (\bibinfo{year}{2019}),
  \urlprefix\url{https://link.aps.org/doi/10.1103/PhysRevLett.122.248004}.

\bibitem[{\citenamefont{Chen et~al.}(2019)\citenamefont{Chen, Bardall, Shearer,
  and Daniels}}]{ChenDanielsSM2019}
\bibinfo{author}{\bibfnamefont{S.-Y.} \bibnamefont{Chen}},
  \bibinfo{author}{\bibfnamefont{A.}~\bibnamefont{Bardall}},
  \bibinfo{author}{\bibfnamefont{M.}~\bibnamefont{Shearer}}, \bibnamefont{and}
  \bibinfo{author}{\bibfnamefont{K.~E.} \bibnamefont{Daniels}},
  \bibinfo{journal}{Soft Matter} \textbf{\bibinfo{volume}{15}},
  \bibinfo{pages}{9426} (\bibinfo{year}{2019}),
  \urlprefix\url{http://dx.doi.org/10.1039/C9SM01756A}.

\bibitem[{\citenamefont{van Gorcum et~al.}(2020)\citenamefont{van Gorcum,
  Karpitschka, Andreotti, and Snoeijer}}]{Gorcum2020}
\bibinfo{author}{\bibfnamefont{M.}~\bibnamefont{van Gorcum}},
  \bibinfo{author}{\bibfnamefont{S.}~\bibnamefont{Karpitschka}},
  \bibinfo{author}{\bibfnamefont{B.}~\bibnamefont{Andreotti}},
  \bibnamefont{and} \bibinfo{author}{\bibfnamefont{J.~H.}
  \bibnamefont{Snoeijer}}, \bibinfo{journal}{Soft Matter}
  \textbf{\bibinfo{volume}{16}}, \bibinfo{pages}{1306} (\bibinfo{year}{2020}),
  \urlprefix\url{http://dx.doi.org/10.1039/C9SM01453E}.

\bibitem[{\citenamefont{Rusanov}(1975)}]{Rusanov:1975aa}
\bibinfo{author}{\bibfnamefont{A.}~\bibnamefont{Rusanov}},
  \bibinfo{journal}{Colloid J. Ussr} \textbf{\bibinfo{volume}{37}},
  \bibinfo{pages}{614} (\bibinfo{year}{1975}).

\bibitem[{\citenamefont{Shanahan}(1987)}]{Shanahan1987aa}
\bibinfo{author}{\bibfnamefont{M.}~\bibnamefont{Shanahan}},
  \bibinfo{journal}{J. Phys. D: Appl. Phys.} \textbf{\bibinfo{volume}{20}},
  \bibinfo{pages}{945} (\bibinfo{year}{1987}).

\bibitem[{\citenamefont{Carr{\'e} et~al.}(1996)\citenamefont{Carr{\'e}, Gastel,
  and Shanahan}}]{Carre1996a}
\bibinfo{author}{\bibfnamefont{A.}~\bibnamefont{Carr{\'e}}},
  \bibinfo{author}{\bibfnamefont{J.-C.} \bibnamefont{Gastel}},
  \bibnamefont{and} \bibinfo{author}{\bibfnamefont{M.~E.~R.}
  \bibnamefont{Shanahan}}, \bibinfo{journal}{Nature}
  \textbf{\bibinfo{volume}{379}}, \bibinfo{pages}{432} (\bibinfo{year}{1996}).

\bibitem[{\citenamefont{White}(2003)}]{White:2003aa}
\bibinfo{author}{\bibfnamefont{L.}~\bibnamefont{White}}, \bibinfo{journal}{J.
  Colloid Interface Sci.} \textbf{\bibinfo{volume}{258}}, \bibinfo{pages}{82}
  (\bibinfo{year}{2003}).

\bibitem[{\citenamefont{Pericet-Camara
  et~al.}(2008)\citenamefont{Pericet-Camara, Best, Butt, and
  Bonaccurso}}]{PC2008aa}
\bibinfo{author}{\bibfnamefont{R.}~\bibnamefont{Pericet-Camara}},
  \bibinfo{author}{\bibfnamefont{A.}~\bibnamefont{Best}},
  \bibinfo{author}{\bibfnamefont{H.~J.} \bibnamefont{Butt}}, \bibnamefont{and}
  \bibinfo{author}{\bibfnamefont{E.}~\bibnamefont{Bonaccurso}},
  \bibinfo{journal}{Langmuir} \textbf{\bibinfo{volume}{24}},
  \bibinfo{pages}{10565} (\bibinfo{year}{2008}).

\bibitem[{\citenamefont{Jerison et~al.}(2011)\citenamefont{Jerison, Xu, Wilen,
  and Dufresne}}]{Jerison2011a}
\bibinfo{author}{\bibfnamefont{E.~R.} \bibnamefont{Jerison}},
  \bibinfo{author}{\bibfnamefont{Y.}~\bibnamefont{Xu}},
  \bibinfo{author}{\bibfnamefont{L.~A.} \bibnamefont{Wilen}}, \bibnamefont{and}
  \bibinfo{author}{\bibfnamefont{E.~R.} \bibnamefont{Dufresne}},
  \bibinfo{journal}{Phys. Rev. Lett.} \textbf{\bibinfo{volume}{106}},
  \bibinfo{pages}{186103} (\bibinfo{year}{2011}).

\bibitem[{\citenamefont{Park et~al.}(2014)\citenamefont{Park, Weon, Lee, Lee,
  Kim, and Je}}]{PARKNATURE}
\bibinfo{author}{\bibfnamefont{S.}~\bibnamefont{Park}},
  \bibinfo{author}{\bibfnamefont{B.}~\bibnamefont{Weon}},
  \bibinfo{author}{\bibfnamefont{J.}~\bibnamefont{Lee}},
  \bibinfo{author}{\bibfnamefont{J.}~\bibnamefont{Lee}},
  \bibinfo{author}{\bibfnamefont{J.}~\bibnamefont{Kim}}, \bibnamefont{and}
  \bibinfo{author}{\bibfnamefont{J.}~\bibnamefont{Je}}, \bibinfo{journal}{Nat
  Commun} \textbf{\bibinfo{volume}{5}}, \bibinfo{pages}{4369}
  (\bibinfo{year}{2014}).

\bibitem[{\citenamefont{Marchand
  et~al.}(2012{\natexlab{b}})\citenamefont{Marchand, Das, Snoeijer, and
  Andreotti}}]{Marchand2012a}
\bibinfo{author}{\bibfnamefont{A.}~\bibnamefont{Marchand}},
  \bibinfo{author}{\bibfnamefont{S.}~\bibnamefont{Das}},
  \bibinfo{author}{\bibfnamefont{J.~H.} \bibnamefont{Snoeijer}},
  \bibnamefont{and}
  \bibinfo{author}{\bibfnamefont{B.}~\bibnamefont{Andreotti}},
  \bibinfo{journal}{Phys. Rev. Lett.} \textbf{\bibinfo{volume}{109}},
  \bibinfo{pages}{236101} (\bibinfo{year}{2012}{\natexlab{b}}).

\bibitem[{\citenamefont{Style and Dufresne}(2012)}]{Style2012a}
\bibinfo{author}{\bibfnamefont{R.~W.} \bibnamefont{Style}} \bibnamefont{and}
  \bibinfo{author}{\bibfnamefont{E.~R.} \bibnamefont{Dufresne}},
  \bibinfo{journal}{Soft Matter} \textbf{\bibinfo{volume}{8}},
  \bibinfo{pages}{7177} (\bibinfo{year}{2012}).

\bibitem[{\citenamefont{Limat}(2012)}]{Limat2012a}
\bibinfo{author}{\bibfnamefont{L.}~\bibnamefont{Limat}}, \bibinfo{journal}{Eur.
  Phys. J. E Soft Matter} \textbf{\bibinfo{volume}{35}}, \bibinfo{pages}{1}
  (\bibinfo{year}{2012}), ISSN \bibinfo{issn}{1292-895X}.

\bibitem[{\citenamefont{Style et~al.}(2013)\citenamefont{Style, Boltyanskiy,
  Che, Wettlaufer, Wilen, and Dufresne}}]{Style2013b}
\bibinfo{author}{\bibfnamefont{R.~W.} \bibnamefont{Style}},
  \bibinfo{author}{\bibfnamefont{R.}~\bibnamefont{Boltyanskiy}},
  \bibinfo{author}{\bibfnamefont{Y.}~\bibnamefont{Che}},
  \bibinfo{author}{\bibfnamefont{J.~S.} \bibnamefont{Wettlaufer}},
  \bibinfo{author}{\bibfnamefont{L.~A.} \bibnamefont{Wilen}}, \bibnamefont{and}
  \bibinfo{author}{\bibfnamefont{E.~R.} \bibnamefont{Dufresne}},
  \bibinfo{journal}{Phys. Rev. Lett.} \textbf{\bibinfo{volume}{110}},
  \bibinfo{pages}{066103} (\bibinfo{year}{2013}).

\bibitem[{\citenamefont{{van Brummelen}
  et~al.}(2017{\natexlab{a}})\citenamefont{{van Brummelen}, Shokrpour~Roudbari,
  Simsek, and {van der Zee}}}]{Brummelen:2017sh}
\bibinfo{author}{\bibfnamefont{E.}~\bibnamefont{{van Brummelen}}},
  \bibinfo{author}{\bibfnamefont{M.}~\bibnamefont{Shokrpour~Roudbari}},
  \bibinfo{author}{\bibfnamefont{G.}~\bibnamefont{Simsek}}, \bibnamefont{and}
  \bibinfo{author}{\bibfnamefont{K.}~\bibnamefont{{van der Zee}}}, in
  \emph{\bibinfo{booktitle}{Fluid Structure Interaction}}, edited by
  \bibinfo{editor}{\bibfnamefont{S.}~\bibnamefont{Frei}},
  \bibinfo{editor}{\bibfnamefont{B.}~\bibnamefont{Holm}},
  \bibinfo{editor}{\bibfnamefont{T.}~\bibnamefont{Richter}},
  \bibinfo{editor}{\bibfnamefont{T.}~\bibnamefont{Wick}}, \bibnamefont{and}
  \bibinfo{editor}{\bibfnamefont{H.}~\bibnamefont{Yang}}
  (\bibinfo{publisher}{De Gruyter}, \bibinfo{year}{2017}{\natexlab{a}}), pp.
  \bibinfo{pages}{283--328}.

\bibitem[{\citenamefont{Lubbers et~al.}(2014)\citenamefont{Lubbers, Weijs,
  Botto, Das, Andreotti, and Snoeijer}}]{LubbersJFM14}
\bibinfo{author}{\bibfnamefont{L.~A.} \bibnamefont{Lubbers}},
  \bibinfo{author}{\bibfnamefont{J.~H.} \bibnamefont{Weijs}},
  \bibinfo{author}{\bibfnamefont{L.}~\bibnamefont{Botto}},
  \bibinfo{author}{\bibfnamefont{S.}~\bibnamefont{Das}},
  \bibinfo{author}{\bibfnamefont{B.}~\bibnamefont{Andreotti}},
  \bibnamefont{and} \bibinfo{author}{\bibfnamefont{J.~H.}
  \bibnamefont{Snoeijer}}, \bibinfo{journal}{J. Fluid Mech. Rapids}
  \textbf{\bibinfo{volume}{747}} (\bibinfo{year}{2014}).

\bibitem[{\citenamefont{Becker and Rannacher}(2001)}]{BeckerRannacher2001}
\bibinfo{author}{\bibfnamefont{R.}~\bibnamefont{Becker}} \bibnamefont{and}
  \bibinfo{author}{\bibfnamefont{R.}~\bibnamefont{Rannacher}},
  \bibinfo{journal}{Acta Numerica} \textbf{\bibinfo{volume}{10}},
  \bibinfo{pages}{1} (\bibinfo{year}{2001}).

\bibitem[{\citenamefont{Oden and Prudhomme}(2001)}]{Oden:2001ss}
\bibinfo{author}{\bibfnamefont{J.}~\bibnamefont{Oden}} \bibnamefont{and}
  \bibinfo{author}{\bibfnamefont{S.}~\bibnamefont{Prudhomme}},
  \bibinfo{journal}{Comput. Math. Appl.} \textbf{\bibinfo{volume}{41}},
  \bibinfo{pages}{735} (\bibinfo{year}{2001}).

\bibitem[{\citenamefont{Nochetto and Veeser}(2012)}]{Nochetto:2012hl}
\bibinfo{author}{\bibfnamefont{R.}~\bibnamefont{Nochetto}} \bibnamefont{and}
  \bibinfo{author}{\bibfnamefont{A.}~\bibnamefont{Veeser}},
  \emph{\bibinfo{title}{Multiscale and Adaptivity: Modeling, Numerics and
  Applications}} (\bibinfo{publisher}{Springer Berlin Heidelberg},
  \bibinfo{year}{2012}), vol. \bibinfo{volume}{2040} of
  \emph{\bibinfo{series}{Lecture Notes in Mathematics}}, chap.
  \bibinfo{chapter}{Primer of Adaptive Finite Element Methods}, pp.
  \bibinfo{pages}{125--225}.

\bibitem[{\citenamefont{{van Brummelen}
  et~al.}(2017{\natexlab{b}})\citenamefont{{van Brummelen}, Zhuk, and {van
  Zwieten}}}]{Brummelen:2017rr}
\bibinfo{author}{\bibfnamefont{E.}~\bibnamefont{{van Brummelen}}},
  \bibinfo{author}{\bibfnamefont{S.}~\bibnamefont{Zhuk}}, \bibnamefont{and}
  \bibinfo{author}{\bibfnamefont{G.}~\bibnamefont{{van Zwieten}}},
  \bibinfo{journal}{Comput. Methods Appl. Mech. Engrg.}
  \textbf{\bibinfo{volume}{313}}, \bibinfo{pages}{723}
  (\bibinfo{year}{2017}{\natexlab{b}}),
  \urlprefix\url{http://www.sciencedirect.com/science/article/pii/S0045782516301815}.

\bibitem[{\citenamefont{{van Zwieten} et~al.}()\citenamefont{{van Zwieten},
  {van Zwieten}, Verhoosel, Fonn, {van Opstal}, and Hoitinga}}]{nutils}
\bibinfo{author}{\bibfnamefont{G.}~\bibnamefont{{van Zwieten}}},
  \bibinfo{author}{\bibfnamefont{J.}~\bibnamefont{{van Zwieten}}},
  \bibinfo{author}{\bibfnamefont{C.}~\bibnamefont{Verhoosel}},
  \bibinfo{author}{\bibfnamefont{E.}~\bibnamefont{Fonn}},
  \bibinfo{author}{\bibfnamefont{T.}~\bibnamefont{{van Opstal}}},
  \bibnamefont{and} \bibinfo{author}{\bibfnamefont{W.}~\bibnamefont{Hoitinga}},
  \urlprefix\url{https://dx.doi.org/10.5281/zenodo.3243447}.

\bibitem[{\citenamefont{Willem}(2013.)}]{Willem2013}
\bibinfo{author}{\bibfnamefont{M.}~\bibnamefont{Willem}},
  \emph{\bibinfo{title}{Functional Analysis}}, Cornerstones,
  (\bibinfo{publisher}{Springer New York :}, \bibinfo{address}{New York, NY :},
  \bibinfo{year}{2013.}),
  \urlprefix\url{http://dx.doi.org/10.1007/978-1-4614-7004-5}.

\bibitem[{\citenamefont{Eggers and Fontelos}(2015)}]{EggersFontelos2015}
\bibinfo{author}{\bibfnamefont{J.}~\bibnamefont{Eggers}} \bibnamefont{and}
  \bibinfo{author}{\bibfnamefont{M.~A.} \bibnamefont{Fontelos}},
  \emph{\bibinfo{title}{Singularities: Formation, Structure, and Propagation}},
  Cambridge Texts in Applied Mathematics (\bibinfo{publisher}{Cambridge
  University Press}, \bibinfo{year}{2015}).

\bibitem[{\citenamefont{Singh and Pipkin}(1965)}]{Singh1965}
\bibinfo{author}{\bibfnamefont{M.}~\bibnamefont{Singh}} \bibnamefont{and}
  \bibinfo{author}{\bibfnamefont{A.~C.} \bibnamefont{Pipkin}},
  \bibinfo{journal}{Zeitschrift für angewandte Mathematik und Physik {ZAMP}}
  \textbf{\bibinfo{volume}{16}}, \bibinfo{pages}{706} (\bibinfo{year}{1965}).

\bibitem[{\citenamefont{Lhermerout et~al.}(2016)\citenamefont{Lhermerout,
  Perrin, Rolley, Andreotti, and Davitt}}]{Lhermerout2016aa}
\bibinfo{author}{\bibfnamefont{R.}~\bibnamefont{Lhermerout}},
  \bibinfo{author}{\bibfnamefont{H.}~\bibnamefont{Perrin}},
  \bibinfo{author}{\bibfnamefont{E.}~\bibnamefont{Rolley}},
  \bibinfo{author}{\bibfnamefont{B.}~\bibnamefont{Andreotti}},
  \bibnamefont{and} \bibinfo{author}{\bibfnamefont{K.}~\bibnamefont{Davitt}},
  \bibinfo{journal}{Nat Commun} \textbf{\bibinfo{volume}{7}},
  \bibinfo{pages}{12545} (\bibinfo{year}{2016}).

\bibitem[{\citenamefont{Bostwick and Daniels}(2013)}]{bostwick2013capillary}
\bibinfo{author}{\bibfnamefont{J.~B.} \bibnamefont{Bostwick}} \bibnamefont{and}
  \bibinfo{author}{\bibfnamefont{K.~E.} \bibnamefont{Daniels}},
  \bibinfo{journal}{Physical Review E} \textbf{\bibinfo{volume}{88}},
  \bibinfo{pages}{042410} (\bibinfo{year}{2013}).

\bibitem[{\citenamefont{Caroli and Nozi{\`e}res}(1998)}]{Caroli1998}
\bibinfo{author}{\bibfnamefont{C.}~\bibnamefont{Caroli}} \bibnamefont{and}
  \bibinfo{author}{\bibfnamefont{P.}~\bibnamefont{Nozi{\`e}res}},
  \bibinfo{journal}{The European Physical Journal B - Condensed Matter and
  Complex Systems} \textbf{\bibinfo{volume}{4}}, \bibinfo{pages}{233}
  (\bibinfo{year}{1998}),
  \urlprefix\url{https://doi.org/10.1007/s100510050374}.

\bibitem[{\citenamefont{Eshelby}(1975)}]{Eshelby1975}
\bibinfo{author}{\bibfnamefont{J.~D.} \bibnamefont{Eshelby}},
  \bibinfo{journal}{Journal of Elasticity} \textbf{\bibinfo{volume}{5}},
  \bibinfo{pages}{321} (\bibinfo{year}{1975}).

\bibitem[{\citenamefont{Rice}(1968)}]{Rice68}
\bibinfo{author}{\bibfnamefont{J.~R.} \bibnamefont{Rice}},
  \bibinfo{journal}{Journal of Applied Mechanics}
  \textbf{\bibinfo{volume}{35}}, \bibinfo{pages}{379} (\bibinfo{year}{1968}),
  ISSN \bibinfo{issn}{0021-8936},
  \eprint{https://asmedigitalcollection.asme.org/appliedmechanics/article-pdf/35/2/379/5449160/379\_1.pdf},
  \urlprefix\url{https://doi.org/10.1115/1.3601206}.

\bibitem[{\citenamefont{de~Gennes et~al.}(2002)\citenamefont{de~Gennes,
  Brochard-Wyart, and Qu{\'e}r{\'e}}}]{deGe02}
\bibinfo{author}{\bibfnamefont{P.-G.} \bibnamefont{de~Gennes}},
  \bibinfo{author}{\bibfnamefont{F.}~\bibnamefont{Brochard-Wyart}},
  \bibnamefont{and}
  \bibinfo{author}{\bibfnamefont{D.}~\bibnamefont{Qu{\'e}r{\'e}}},
  \emph{\bibinfo{title}{Capillarity and Wetting Phenomena: Drops, Bubbles,
  Pearls, Waves}} (\bibinfo{publisher}{Belin}, \bibinfo{year}{2002}).

\bibitem[{\citenamefont{van Gorcum et~al.}(2018)\citenamefont{van Gorcum,
  Andreotti, Snoeijer, and Karpitschka}}]{Gorcum2018}
\bibinfo{author}{\bibfnamefont{M.}~\bibnamefont{van Gorcum}},
  \bibinfo{author}{\bibfnamefont{B.}~\bibnamefont{Andreotti}},
  \bibinfo{author}{\bibfnamefont{J.~H.} \bibnamefont{Snoeijer}},
  \bibnamefont{and}
  \bibinfo{author}{\bibfnamefont{S.}~\bibnamefont{Karpitschka}},
  \bibinfo{journal}{Phys. Rev. Lett.} \textbf{\bibinfo{volume}{121}},
  \bibinfo{pages}{208003} (\bibinfo{year}{2018}),
  \urlprefix\url{https://link.aps.org/doi/10.1103/PhysRevLett.121.208003}.

\end{thebibliography}
\end{document}